\documentclass[11pt,a4paper]{article}

\usepackage[margin=1in]{geometry}
\usepackage{setspace}
\usepackage{lineno}
\usepackage[T1]{fontenc}
\usepackage[utf8]{inputenc}
\usepackage{lmodern}
\usepackage{amsmath,amssymb,amsfonts}
\usepackage{graphicx}
\usepackage{placeins}
\usepackage{tikz}
\usetikzlibrary{shapes,arrows.meta,positioning,calc}
\usepackage{booktabs,longtable}
\usepackage{caption}
\usepackage{xurl}
\usepackage[authoryear,round]{natbib}
\bibpunct{(}{)}{;}{a}{}{,}
\usepackage{hyperref}
\hypersetup{colorlinks=true,allcolors=black}
\usepackage{ragged2e}
\justifying
\providecommand{\Description}[1]{}

\begin{document}

\begin{center}
{\Large\bfseries Migration Genealogies in Multiregional Stable Populations: First-Return Decompositions, Recurrence, and Sensitivity}
\end{center}
\begin{center}
Ryo Oizumi\textsuperscript{1,*}, Kensaku Kinjo\textsuperscript{2}, Yuki Chino\textsuperscript{3}
\end{center}
\begingroup\small\singlespacing
\noindent\textsuperscript{1}International Relations, National Institute of Population and Social Security Research, 2-2-3 Uchisaiwai-cho, Chiyoda-ku, Tokyo 100-0011, Japan.\par
\noindent\textsuperscript{2}Kushiro college, National Institute of Technology, Otanoshike-Nishi 2-32-1, Kushiro-Shi, Hokkaido 084-0916, Japan. Email: \url{kinjo@kushiro-ct.ac.jp}.\par
\noindent\textsuperscript{3}Department of Applied Mathematics, National Yang Ming Chiao Tung University, Hsinchu 30010, Taiwan. Email: \url{chino@nycu.edu.tw}.\par
\medskip
\noindent\textsuperscript{*}Corresponding author: Ryo Oizumi, International Relations, National Institute of Population and Social Security Research, 2-2-3 Uchisaiwai-cho, Chiyoda-ku, Tokyo 100-0011, Japan. Email: \url{ooizumi-ryou@ipss.go.jp}.\par
\endgroup

\begin{center}\small Preprint\end{center}
\thispagestyle{plain}
\setcounter{page}{1}

\section*{Abstract}

When people of reproductive age move between regions, replacement-level fertility need not produce spatially balanced population replacement. Migration genealogies connect a woman's birth region, through survival and movement, to her daughters' birth regions across generations. We decompose existing genealogical series for stable births and reproductive value into first-return paths, on which a lineage again produces a descendant in its reference region without intermediate births there. Discounting at the intrinsic growth rate yields a first-return distribution with an Euler--Lotka-type normalization. Its mean recurrence generation, the expected generations until first return, equals the normalized product of stable births and reproductive value and, at replacement, reciprocal net reproduction sensitivity to within-region reproduction. Using Japan's 2010, 2015, and 2020 censuses, vital statistics, and life tables, we compare 47 prefectures under national replacement. Based on 2020 rates, mean recurrence is approximately 10 generations for Tokyo and 500 for Akita; Fukushima varies across observation years. Similar one-generation birth contributions conceal different descendant destinations and subsequent return pathways. These period-based results reveal how reproductive pathways generate regional differences in stationary populations. Population replacement must therefore be analyzed jointly through fertility, migration, and regional conditions, with mean recurrence identifying the strength of closed reproductive pathways.

\noindent\textbf{Keywords:} multiregional stable population; migration genealogy; first return; reproductive value; population growth sensitivity

\clearpage

\section*{Introduction}

How do migration and the subsequent births of descendants shape the regional population distribution when national reproduction is held at replacement? By changing where a woman gives birth, migration also changes the region in which the next generation begins its life course. Its demographic effects can therefore extend beyond the redistribution of existing residents to the regional pathways through which their descendants contribute future births. Age-structured multiregional population theory describes these processes through fertility, survival, and movement among regions. The intrinsic growth rate, stable age distribution, and reproductive value describe long-run population change. For the multiregional Leslie matrix, the Perron root is the population growth multiplier $\lambda=e^r$; the intrinsic growth rate is $r=\log\lambda$, expressed per five-year period, with annual equivalent $r/5$. The right and left eigenvectors represent the stable age distribution across regions and reproductive value. The Perron root of the undiscounted next-generation matrix instead gives the net reproduction rate. The next-generation matrix connects these age-specific processes to reproduction between birth regions. These objects are central to the multiregional and multigroup foundations of \citet{Rogers1975} and \citet{Schoen1988}, the matrix population and sensitivity framework of \citet{Caswell2001}, and the multiregional and operator formulations of \citet{Inaba1986,Inaba2009,Inaba2017}.

Classical multiregional theory determines the stable regional distribution for a given demographic operator and anticipates that national replacement need not imply uniform regional replacement. The stable age distribution and reproductive value, however, compress many demographic paths into a small number of quantities. Migration operates on two time scales: movement experienced by individuals over the life course and migration genealogies transmitted across generations. An entry of the next-generation matrix aggregates a woman's survival and movement from her birth region to the birth of a daughter in a destination region; powers of that matrix connect these mother-to-daughter contributions across successive generations. These demographic summaries do not, by themselves, display how these paths leave a birth region, pass through other regions, and return through the births of descendants. Because stable age distributions and reproductive values sum contributions over generations, they do not directly distinguish the generation of first return.

\citet{Oizumi2022PLOS} derived matrix-entry series for the stable age distribution and reproductive value and interpreted their terms through individual and ancestral migration histories. They combined these representations with population growth sensitivity analysis and applied them to Japanese prefectural data. We build on those series, including paths that avoid a reference region at intermediate steps. Such paths trace lineages whose intermediate descendant births occur outside the reference region before a descendant is again born there.

To explain the intergenerational reproductive mechanisms determining the stationary population distribution under national replacement, we decompose reproduction into recurrence pathways. We ask whether the inherited reference-avoiding terms define a normalized first-return probability law when discounted at the intrinsic growth rate and how its mean relates to stable births, reproductive value, and local reproductive sensitivity. We interpret the inherited characteristic-equation relation as an Euler--Lotka-type normalization of the first-return distribution under discounting at the intrinsic growth rate, define its mean recurrence generation as the expected number of generations until first return, and prove the identity connecting this mean to the normalized product of stable births and reproductive value and, at replacement, reciprocal sensitivity of net reproduction to within-region reproduction. Together, the recurrence probability structure and these identities identify and quantify the first-return mechanisms underlying the stable age distribution and reproductive value. Under a common national replacement condition, they reveal how different reproductive pathways generate regional unevenness even when fertility or one-generation contributions are similar.

In the inherited series, the stable distribution of births identifies descendant destinations, whereas reproductive value identifies contributions toward the reference region. Appendix B distinguishes the inherited series and characteristic-equation representation from Proposition B.1 and the recurrence--sensitivity identity derived here.

Mean recurrence is a structural quantity of the operator defined by period fertility, survival, and migration rates, just as stable population, intrinsic growth, reproductive value, and generation time are period-model quantities. A large mean identifies weak closed reproductive pathways to the reference region under that operator. It does not assume that actual behavior will remain unchanged for hundreds of generations. Comparison across observation dates measures changes in the operators and their recurrence structures.

Japan's 47 prefectures provide a spatially detailed application with comparable population, fertility, mortality, and migration information for 2010, 2015, and 2020. These data allow comparison of regional reproductive pathways formed through internal migration and births. We use the population censuses \citep{Census2010,Census2015,Census2020}, vital statistics \citep{MHLWVitalStatistics}, and prefectural life tables \citep{MHLWLifeTable2010,MHLWLifeTable2015,MHLWLifeTable2020}. Common fertility scaling sets national net reproduction to one while retaining relative prefectural differences. This benchmark permits comparison of genealogical structure under a common replacement condition. Extending the data to 2020 supports the empirical comparison; the contribution to demographic theory is the recurrence law and its identities.

Akita is a prefecture in northeastern Japan that has experienced outmigration, whereas Tokyo is the core of the capital metropolitan region, a major destination for internal migration. As a hypothetical genealogy illustrating first return, consider a woman born in Akita who moves to Tokyo and gives birth to a daughter there. Her move belongs to an individual life course; her daughter's birthplace begins the next step in a migration genealogy. If that daughter later gives birth in Akita, this line of descent first returns to Akita in the second reproductive generation. If births continue elsewhere, its first return occurs later. Return here concerns a descendant's birthplace, not the original woman's return migration.

The Japanese application shows why this distinction matters. Tokyo, Akita, and Kyoto, a prefecture in western Japan, have broadly similar, low one-generation contributions to births within their own prefectures, yet their mean recurrence generations differ markedly. A daughter born elsewhere may enter pathways that contribute births back to the reference prefecture or pathways concentrated in other regions. Distinguishing the generation of first return reveals how these alternatives contribute to the stationary population distribution, beyond the total contribution in one generation or the stable distribution that sums across generations.

Internal-migration research documents different spatial trajectories of population decline \citep{Rowe2019,Inoue2022,NewshamRowe2023}, age-specific movement through Japan's urban hierarchy \citep{KotsuboNakaya2024}, and changes around the pandemic \citep{KotsuboNakaya2023}. Disaster-related movements also follow distinct migration systems \citep{Hauer2020}. These findings motivate comparisons of period operators, particularly for Fukushima. More broadly, the dependence of population scenarios on demographic assumptions \citep{Vollset2020} makes it essential to distinguish structural period measures from forecasts. The Japanese application quantifies reproductive pathways from observed fertility, survival, and migration.

\section*{Methods}

\subsection*{Model Framework}

We represent Japan's female population as a \emph{multistate age-structured population} in which fertility, survival, and migration jointly transmit regional residence through reproduction. Two time scales must be distinguished: $K_{ij}(a\mid s)$ describes one woman's survival and movement within a lifetime, whereas $\pi_{ij}^{j}(m;r)$ and its adjoint describe genealogical paths across $m$ reproductive generations. The multiregional Leslie matrix, genealogical series for the stable age distribution and reproductive value, and sensitivity framework build on \citet{Oizumi2022PLOS}. Here we use the first-return terms of those series to establish the probability law and identities connecting recurrence to stable births, reproductive value, and diagonal sensitivity.

Table~\ref{tab:notation} summarizes the principal notation used for the multiregional model, migration genealogies, and recurrence measures.

Let $n_t(a,i)$ denote the cohort of women in age class $a$ residing in prefecture $i$ at time $t$ in a system of $M$ prefectures. Let $T_{ij}(a)$ be the probability that a woman in age class $a$ moves from prefecture $j$ to prefecture $i$, satisfying
\begin{equation}
\sum_{i=1}^{M}T_{ij}(a)=1,\quad
T_{jj}(a)=1-\sum_{i\neq j}^{M}T_{ij}(a),\qquad
0\leq T_{ij}(a)\leq 1 \quad \text{for all }a.
\end{equation}
Let $p_i(a)$ denote survival in prefecture $i$ over $[a,a+1]$. One unit of time and age in the model represents five years, so $a$ indexes a five-year cohort. Cohort migration is then
\begin{equation}\label{mp1}
n_t(a,i)=\sum_{j=1}^{M}k_{ij}(a-1)n_{t-1}(a-1,j),\qquad
k_{ij}(a):=T_{ij}(a)p_j(a).
\end{equation}
We call $k_{ij}(a)$ the \emph{multiregional survival rate}; it is the joint probability of surviving the age interval and residing in destination region $i$, given residence in region $j$ at its start.

Because birth and migration may in general occur together, the birth process can be written using cross-regional fertility $f_{ij}(a)\geq0$:
\begin{equation}\label{rp1}
n_t(0,i)=\sum_{a=0}^{\omega-1}\sum_{j=1}^{M}f_{ij}(a)n_{t-1}(a,j),
\end{equation}
where $\omega$ denotes the limiting age class. Empirically, age 85 and older is an aggregated terminal class. Transitions into this class are included, but no retention transition from age 85 and older to the next period is specified; members leave the model during the following five years.

Combining age- and region-specific fertility, survival, and migration yields the multiregional Leslie matrix $\mathbf M$. Its first block row contains the birth process in Eq.~\eqref{rp1}, and its subdiagonal blocks contain aging and migration in Eq.~\eqref{mp1}. The population vector evolves according to
\begin{equation}
\mathbf n(t+1)=\mathbf M\mathbf n(t).
\end{equation}
The population growth multiplier is $\lambda=e^r$, where $r$ is the intrinsic growth rate per five-year period. The stable age distribution across prefectures is $\mathbf w$, and reproductive value is $\mathbf v$.

In classical stable population theory, the stable age distribution and reproductive value summarize one-generation contributions through quantities such as expected births, life expectancy, and mean generation time \citep{Keyfitz1968,Caswell2001,Preston2001}. In a multiregional society, however, birthplace, residence, migration, and reproduction are linked across generations. Where a child later reproduces depends on the parent's life course, which was itself shaped by the regional trajectory of earlier generations. Interregional migration contains both individual movement over a lifetime and the return or persistence of a migration genealogy across generations. Within this genealogical framework, we quantify return across generations through the first-return distribution and its mean.

\begin{table}[p]
    \centering
    \caption{Summary of key mathematical notation and demographic interpretation}
    \label{tab:notation}
    \begingroup
    \scriptsize
    \setlength{\tabcolsep}{3pt}
    \renewcommand{\arraystretch}{0.82}
    \begin{tabular}{p{0.16\textwidth}p{0.32\textwidth}p{0.42\textwidth}}
        \toprule
        Notation & Mathematical definition or concept & Demographic interpretation \\
        \midrule
        \multicolumn{3}{l}{\textit{Demographic and spatial variables}} \\
        $M$ & Number of prefectures or spatial regions & Number of regions in the multiregional system; $M=47$ for Japan's prefectures. \\
        $a$ & Five-year age class & Age is indexed in five-year units from $0$--$4$ through the terminal class. \\
        $\omega$ & Number of modeled age classes in the Leslie matrix & Upper limit of the age classes included in the state vector. \\
        $T_{ij}(a)$ & Five-year migration probability from region $j$ to region $i$ at age class $a$ & Probability that a woman in region $j$ moves to region $i$ during the five-year interval. \\
        $p_j(a)$ & Five-year survival probability in region $j$ at age class $a$ & Probability of surviving the age interval, evaluated by region of residence at the beginning of the interval. \\
        $k_{ij}(a)=T_{ij}(a)p_j(a)$ & Multiregional survival rate & Joint probability of surviving the age interval and residing in region $i$, given initial residence in region $j$. \\
        $K_{ij}(a\mid s)$ & Multiregional survivorship from age $s$ in region $j$ to age $a$ in region $i$ & Life-course probability that a woman observed at age $s$ in region $j$ survives and resides in region $i$ at age $a$. \\
        $f_{ij}(a)$ & Cross-regional fertility term & Expected number of female births in region $i$ to women age $a$ associated with region $j$; in the empirical model newborns inherit the mother's region of residence. \\
        \addlinespace
        \multicolumn{3}{l}{\textit{Matrix notation and eigen-quantities}} \\
        $\mathbf M$ & Multiregional Leslie matrix & Age-by-region projection matrix combining fertility, survival, and internal migration. \\
        $\boldsymbol\Psi(r)$ & Discounted next-generation matrix & Matrix of one-generation expected discounted female births by birth region. \\
        $\psi_{ij}(r)$ & Element $(i,j)$ of $\boldsymbol\Psi(r)$ & Expected discounted daughters born in region $i$ to women born in region $j$. \\
        $\rho(\boldsymbol\Psi(0))$ & Dominant eigenvalue of the undiscounted next-generation matrix & Net reproduction rate of the multiregional population before replacement-level normalization. \\
        $\mathbf w,\mathbf v$ & Right and left Perron eigenvectors & $\mathbf w$ gives the stable age-region distribution; $\mathbf v$ gives reproductive value. \\
        \addlinespace
        \multicolumn{3}{l}{\textit{Genealogical and recurrence quantities}} \\
        $\pi_{ij}^{j}(m;r)$ & Weight of $m$-generation genealogical paths from reference region $j$ to region $i$ & Weighted contribution to descendants in region $i$ after $m$ generations, avoiding $j$ at intermediate generations. \\
        $\pi_{jj}^{j}(m;r)$ & First-return weight to reference region $j$ in generation $m$ & First-return path weight; evaluated at the intrinsic growth rate, these weights form a probability distribution over $m$. \\
        $E_j[m]$ & Mean recurrence generation, $\sum_{m\geq1}m\pi_{jj}^{j}(m;r)$ & Expected number of generations until a lineage associated with region $j$ again produces descendants in region $j$. \\
        $\partial\rho/\partial\psi_{jj}$ & Perron-root sensitivity, $\partial\rho(\boldsymbol\Psi(r))/\partial\psi_{jj}(r)$ & Local sensitivity of the dominant eigenvalue to within-region genealogical reproduction in region $j$; its reciprocal is $E_j[m]$ under Perron-root normalization. \\
        \bottomrule
    \end{tabular}
    \endgroup
    \begin{flushleft}
    \footnotesize
    Note: One model period is five years. A generation in the recurrence quantities is a genealogical step through the next-generation matrix, not a forecast horizon in calendar time.
    \end{flushleft}
\end{table}

\FloatBarrier

\subsection*{Measures}

Solving Eq.~\eqref{mp1} recursively decomposes a cohort into its initial state and the distribution of its subsequent life course:
\begin{equation}\label{mp2}
n_t(a,i)=\begin{cases}
\sum_{j=1}^{M}K_{ij}(a\mid a-t)n_0(a-t,j),&a-t\geq0,\\
\sum_{j=1}^{M}K_{ij}(a\mid0)n_{t-a}(0,j),&t-a>0.
\end{cases}
\end{equation}
The probability $K_{ij}(a\mid s)$, the \emph{multiregional survivorship}, is
\begin{equation}
K_{ij}(a\mid s):=\begin{cases}
\displaystyle\sum_{j_1,\ldots,j_{a-s-1}=1}^{M}k_{ij_1}(a-1)k_{j_1j_2}(a-2)\cdots k_{j_{a-s-1}j}(s),&a-1>s,\\
k_{ij}(a-1),&a-1=s,\\
\delta_{ij},&a=s,\\
0,&a<s,
\end{cases}\label{dk1}
\end{equation}
where $\delta_{ij}$ is the Kronecker delta. It is the probability that a woman residing in prefecture $j$ at age $s$ survives and resides in prefecture $i$ at age $a$, summed over all possible migration paths.

Aggregating all age-specific paths from region $j$ at birth through survival and migration to reproduction in region $i$ gives
\begin{equation}
\psi_{ij}(0):=\sum_{a=0}^{\omega-1}\sum_{\ell=1}^{M}f_{i\ell}(a)K_{\ell j}(a\mid0).
\end{equation}
The matrix $\boldsymbol\psi(0):=(\psi_{ij}(0))$ is the \emph{next-generation matrix}; it gives the expected reproduction of an entire birth cohort into the next generation. Its spectral radius $\rho(\boldsymbol\psi)$ is the system's net reproduction rate \citep{Inaba2009}.

Under the convergence conditions stated in Appendix B, long-run population change is calculated from the intrinsic growth rate $r$, reproductive value $\mathbf v=(v(a,i))$, and stable age distribution $\mathbf w=(w(a,i))$ as
\begin{equation}
n_t(a,i)=\frac{\mathbf v\mathbf n_0}{\mathbf v\mathbf w}e^{rt}w(a,i)+O(e^{-\delta t}),\qquad \delta>0.
\end{equation}
The stable-population balance equation $\lambda\mathbf w=\mathbf M\mathbf w$ implies
\begin{align}
\lambda w(a,i)&=\sum_{j=1}^{M}k_{ij}(a-1)w(a-1,j),\\
\lambda w(0,i)&=\sum_{a=0}^{\omega-1}\sum_{j=1}^{M}f_{ij}(a)w(a,j).
\end{align}
Solving the first equation recursively yields
\begin{equation}\label{wa1}
w(a,i)=\sum_{j=1}^{M}\lambda^{-a}K_{ij}(a\mid0)w(0,j).
\end{equation}
Substitution into the equation for $w(0,i)$ gives (see Appendix B)
\begin{align}
w(0,i)&=\begin{cases}c_j\displaystyle\sum_{m=1}^{\infty}\pi_{ij}^{j}(m;r),&i\neq j,\\ c_j,&i=j,\end{cases}\quad c_j\neq0,\label{w1}\\
\pi_{ij}^{j}(1;r)&:=\psi_{ij}(r),\\
\pi_{ij}^{j}(m;r)&:=\sum_{j_1,\ldots,j_{m-1}\neq j}^{M}\psi_{ij_1}(r)\psi_{j_1j_2}(r)\cdots\psi_{j_{m-1}j}(r),\label{pi}\\
\psi_{ij}(r)&:=\sum_{a=0}^{\omega-1}\lambda^{-a-1}\sum_{\ell=1}^{M}f_{i\ell}(a)K_{\ell j}(a\mid0).\label{psi}
\end{align}
Equation~\eqref{psi} gives the expected discounted daughters born in prefecture $i$ to women born in prefecture $j$, after aggregating the women's survival and movement before reproduction. The factor $\lambda^{-a-1}=e^{-r(a+1)}$ discounts the timing of births in five-year model units at the intrinsic growth rate of the same Leslie matrix. In contrast, $m$ in Eq.~\eqref{pi} counts mother-to-daughter steps, not five-year intervals. For $i=j$, the series includes only paths whose intermediate descendant births are outside $j$; the first descendant birth back in $j$ occurs at step $m$. The case $m=1$ includes daughters born directly in $j$, whether or not their mothers moved during their lives. The reference-region balance equation yields
\begin{equation}
\sum_{m=1}^{\infty}\pi_{jj}^{j}(m;r)=1.
\end{equation}
After setting the contribution to the reference prefecture $j$ equal to one (or $c_j$), the stable distribution at age zero describes the multigenerational contribution to all other prefectures.

Reproductive value $\mathbf v$, satisfying $\lambda\mathbf v=\mathbf v\mathbf M$, obeys
\begin{align}
\lambda v(a,i)&=\sum_{j=1}^{M}v(0,j)f_{ji}(a)+\sum_{j=1}^{M}v(a+1,j)k_{ji}(a),\\
v(a,i)&=\sum_{j=1}^{M}v(0,j)\sum_{x=a}^{\omega-1}\lambda^{-(x-a)-1}\sum_{\ell=1}^{M}f_{j\ell}(x)K_{\ell i}(x\mid a).\label{va1}
\end{align}
As with $w(0,i)$, the age-zero values can be expanded as (Appendix B)
\begin{align}
v(0,i)&=\begin{cases}d_j\displaystyle\sum_{m=1}^{\infty}\pi_{ji}^{*j}(m;r),&i\neq j,\\ d_j,&i=j,\end{cases}\quad d_j\neq0,\label{v1}\\
\pi_{ji}^{*j}(1;r)&:=\psi_{ji}(r),\\
\pi_{ji}^{*j}(m;r)&:=\sum_{j_1,\ldots,j_{m-1}\neq j}^{M}\psi_{jj_1}(r)\psi_{j_1j_2}(r)\cdots\psi_{j_{m-1}i}(r).\label{api}
\end{align}
Because the reference prefecture appears on the descendant side in Eq.~\eqref{api}, age-zero reproductive value describes contributions from other origins after the influence of a woman originating in $j$ on descendants originating in $j$ is set to one (or $d_j$). The corresponding reference-region balance equation gives
\begin{equation}\label{ele}
\sum_{m=1}^{\infty}\pi_{jj}^{*j}(m;r)=\sum_{m=1}^{\infty}\pi_{jj}^{j}(m;r)=1.
\end{equation}
Equation~\eqref{ele} is an Euler--Lotka-type normalization over first-return generations: discounted path weights sum to one at the intrinsic growth rate. Setting the reference-region components of $\mathbf v$ and $\mathbf w$ to one fixes their scale; it does not create this probability normalization. The demographic relationships involve two time scales: Eq.~\eqref{dk1} describes the distribution of a single life course, whereas Eq.~\eqref{pi} and \eqref{api} describe migration across generations.

\subsection*{Analytic Strategy}

Because the multiregional Leslie matrix has demographically interpretable stable distributions and reproductive values, we investigate how migration affects population structure and intrinsic growth through stationary profiles:
\begin{equation}\label{sp1}
\lim_{t\to\infty}e^{-rt}n_t(a,i)=\frac{\mathbf v\mathbf n_0}{\mathbf v\mathbf w}w(a,i).
\end{equation}
To evaluate regional characteristics, including migration, while holding national reproduction at replacement level, we divide all multiregional fertility rates by the net reproduction rate, using $f_{ij}(a)/\rho(\boldsymbol\psi)$. This common scaling preserves relative regional fertility differences.

Panel a of Figure~\ref{fig:selfreplace} reports the prefectural female fertility index after this normalization:
\begin{equation}\label{rffi}
I_i^{\mathrm F}:=\sum_{a=0}^{\omega-1}f_i^{\mathrm{rep}}(a)
=\frac{1}{\rho(\boldsymbol\psi(0))}\sum_{a=0}^{\omega-1}f_i(a).
\end{equation}
The numerator $f_i(a)$ is the number of female births per woman, incorporating the female share of births and infant survival. $I_i^{\mathrm F}$ is not the conventionally reported total fertility rate (TFR) for births of both sexes. The reference value $I_i^{\mathrm F}=1$ indicates that the sum of normalized female fertility equals one; it does not mean that prefecture $i$ independently achieves replacement when migration is included.

Panel c of Figure~\ref{fig:age-distribution-reproductive-value} uses the absolute sensitivity of the population growth multiplier $\lambda$ to a marginal change in a matrix element. Here $\lambda$ is the population growth multiplier over one five-year period. For the stable age distribution $\mathbf w$ and reproductive value $\mathbf v$, sensitivities to multiregional survival and diagonal fertility are
\begin{align}
S_{ij}^{K}(a)&:=\frac{\partial\lambda}{\partial k_{ij}(a)}
=\frac{v(a+1,i)w(a,j)}{\mathbf v\mathbf w},\label{sensK}\\
S_i^{F}(a)&:=\frac{\partial\lambda}{\partial f_{ii}(a)}
=\frac{v(0,i)w(a,i)}{\mathbf v\mathbf w}.\label{sensF}
\end{align}
Fertility sensitivity is limited to diagonal entries because implementation assumes $f_{ij}(a)=f_i(a)\delta_{ij}$. For both observed 2020 rates and replacement-level rates, we compare $S_{ij}^{K}(a)$ over every origin $j$ and destination $i$ within each age class and extract the maximum. We likewise extract the maximum fertility sensitivity across prefectures and display values for Tokyo and Aichi. These quantities are absolute sensitivities, not elasticities. Because $k_{ij}(a)$ is varied independently without imposing compensating changes that preserve the column sums of migration probabilities, they are local demographic sensitivity measures rather than effects of feasible migration policies.

Stationary profiles provide useful approximations to the population structures implied by persistent demographic conditions, such as prolonged low fertility. Classical Leslie and McKendrick models at replacement level naturally yield measures such as generations per lifetime and expected births implied by the number and age structure of women ages 0--49. We first expand the term $\mathbf v\mathbf n_0$ in Eq.~\eqref{sp1}:
\begin{equation}\label{I}
\mathbf v\mathbf n_0=
\sum_{m=1}^{\infty}\sum_{a=0}^{\omega-1}\sum_{i,k,\ell=1}^{M}
\pi_{jk}^{*j}(m;r)
\sum_{x=a}^{\omega-1}\lambda^{-(x-a)-1}f_{k\ell}(x)K_{\ell i}(x\mid a)n_0(a,i).
\end{equation}
Arbitrary constants cancel from the stationary profile, so for a reference prefecture $j$ we set $d_j=c_j=1$, or $v(0,j)=w(0,j)=1$. Together with the interpretation of reproductive value, Eq.~\eqref{I} measures the contribution of daughters in the initial population to the reproduction of women originating in $j$ over all generations. At replacement level ($r=0$ and $\lambda=1$), it is the total number of first descendants that daughters of the initial population produce in prefecture $j$. To align the multiregional model with conventional stable population measures, define
\begin{equation}\label{em}
E_j[m]:=\frac{\mathbf v(0)\mathbf w(0)}{v(0,j)w(0,j)}
=\sum_{m=1}^{\infty}m\pi_{jj}^{j}(m;r).
\end{equation}
Appendix B provides the proof. Under $v(0,j)=w(0,j)=1$, the left side is $\sum_i v(0,i)w(0,i)$. Equation~\eqref{em} defines the \emph{mean recurrence generation} as the expected first-return generation under the normalized path weights. It counts steps along a line of descent, rather than the earliest return among all branches of an actual woman's family. At replacement, $r=0$ and the age discount disappears. The quantity remains a measure of the fixed reproductive network, distinct from the calendar duration of a generation. We normalize Eq.~\eqref{I} by this quantity:
\begin{equation}
R_j:=\frac{\mathbf v\mathbf n_0}{E_j[m]}.
\end{equation}
Then $R_j$ is the contribution of the initial population to the reproduction of women originating in $j$ per generation of daughters, consistent with a conventional one-generation measure.

Similarly, normalizing $\mathbf v\mathbf w$ by $E_j[m]$ yields the mean generation time for females:
\begin{equation}\label{tj}
T:=\frac{\mathbf v\mathbf w}{E_j[m]}
=\frac{1}{E_j[m]}\sum_{\nu=1}^{\infty}\sum_{\ell=1}^{\nu-1}\sum_{k,i=1}^{M}
\pi_{jk}^{*j}(\nu-\ell;r)
\left(-\frac{d}{dq}\psi_{ki}(q)\Bigr|_{q=r}\right)
\pi_{ij}^{j}(\ell;r).
\end{equation}
The quantity $T$ is invariant to the choice of $j$.

Lotka's life expectancy for women originating in prefecture $j$ is
\begin{equation}
e_r(j)=\sum_{a=0}^{\omega-1}\sum_{i=1}^{M}\lambda^{-a}K_{ij}(a\mid0).
\end{equation}
The total stable population satisfies
\begin{equation}
\|\mathbf w\|:=\sum_{a=0}^{\omega-1}\sum_{i=1}^{M}w(a,i)
=\sum_{j=1}^{M}e_r(j)w(0,j).
\end{equation}
Normalizing by $\|\mathbf w(0)\|_j:=\sum_i w(0,i)$ gives the invariant Lotka life expectancy
\begin{equation}
e_r:=\frac{\|\mathbf w\|}{\|\mathbf w(0)\|_j}.
\end{equation}
At replacement level, $e_r$ is life expectancy $e_0$. The normalizing constant $\|\mathbf w(0)\|_j$ represents the contribution of women originating in $j$ to reproduction in all regions. Because one model period is five years, $T$ and $e_r$ are multiplied by five when reported in years.

Finally, let
\begin{equation}
F_0:=\|\mathbf w(0)\|_jR_j.
\end{equation}
This is the contribution of descendants of the initial population to all regions after passing through prefecture $j$ in some generation, and it is invariant to $j$. We assume the next-generation matrix is irreducible: for any reference prefecture, a descendant can appear there within a finite number of generations. In the Japanese data the matrix is usually positive, meaning a nonzero probability of descendants in every prefecture within one generation. The total stationary female population at replacement level retains the classical form: expected births $F_0$ times the number of generations per lifetime, $e_0/T$:
\begin{equation}\label{total}
\sum_{a=0}^{\omega-1}\sum_{i=1}^{M}
\frac{\mathbf v\mathbf n_0}{\mathbf v\mathbf w}w(a,i)
=\frac{F_0e_r}{T}.
\end{equation}
For $\mathbf n_0$ we use the female population by age and prefecture in each census year. Because reproductive value is zero at ages 50 and older, only women ages 0--49 contribute to $\mathbf v\mathbf n_0$. We apply the same stable-population projection to the nationally aggregated initial female population in the classical single-region model. In the multiregional model, the stationary female population is the sum of Eq.~\eqref{total} over all ages and 47 prefectures. The constants $E_j[m]$ and $\|\mathbf w(0)\|_j$ remain important measures of regional structure and distinguish population-receiving from population-losing regions.

The key objects are the first-return distribution $\pi_{jj}^{j}(m;r)$, which determines $E_j[m]$, and the path weights $\pi_{ij}^{j}(m;r)$, which quantify contributions to descendants in region $i$ from region $j$ through paths avoiding $j$ at intermediate generations. These are constructed from the one-generation reproductive contributions $\psi_{ij}$. Within-generation life courses are given statistically by the multiregional survivorship in Eq.~\eqref{dk1}.

For panel b of Figure~\ref{fig:life-course-migration}, we fix terminal age $a=85+$ and terminal prefecture $i$ and vary age $s$ and residence $j$ to define the \emph{backward profile}
\begin{equation}
B_i(s,j):=K_{ij}(85+\mid s).
\label{backwardprofile}
\end{equation}
This is the probability that a woman residing in prefecture $j$ at age $s$ survives and resides in prefecture $i$ at age 85 or older. We set $B_i(85+,j)=\delta_{ij}$ and display the probability in Eq.~\eqref{backwardprofile} directly.

Using the 2010, 2015, and 2020 censuses, which report residence five years earlier by age \citep{Census2010,Census2015,Census2020}, together with other government statistics, we calculate stationary profiles and the stationary populations implied if replacement level were reached after each census year. We contrast population-receiving and population-losing regions and simulate multiregional survivorship and the regional origins of descendants by generation. Appendix A documents data sources, transformations, and additional assumptions.

\subsection*{Focus of the Analysis}

We focus on the relationship between the mean recurrence generation in Eq.~\eqref{em} and the stationary population of reference prefecture $j$. For the stable distribution of births $\mathbf w(0)$ and reproductive value at birth $\mathbf v(0)$, evaluated at the intrinsic growth rate, their genealogical interpretations are expressed by
\begin{align}
\frac{\mathbf w(0)}{w(0,j)}&=
\left(\sum_{m=1}^{\infty}\pi_{ij}^{j}(m;r)\right)_{1\leq i\leq M}^{\top},\\
\frac{\mathbf v(0)}{v(0,j)}&=
\left(\sum_{m=1}^{\infty}\pi_{ji}^{*j}(m;r)\right)_{1\leq i\leq M}.
\end{align}
Define
\begin{equation}
\kappa_j:=\frac{w(0,j)v(0,j)}{\mathbf v(0)\mathbf w(0)}.
\end{equation}
Then
\begin{equation}
E_j[m]=\kappa_j^{-1}=\frac{\mathbf v(0)\mathbf w(0)}{w(0,j)v(0,j)}.
\end{equation}
The quantity $\kappa_j$ is invariant to arbitrary scaling of the stable age distribution and reproductive value and depends only on $j$. It is small where both the age-zero population and reproductive value are small relative to other regions, and the mean recurrence generation is correspondingly large. Where descendants of women who leave require many generations to return, the prefecture has a small relative population, low reproductive value, or both, even under replacement fertility.

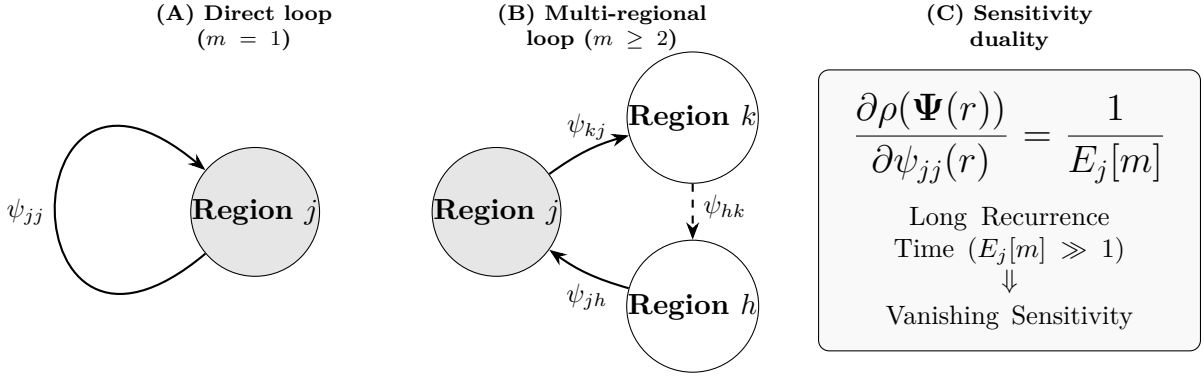
\begin{figure}[!htbp]
\centering
\resizebox{\textwidth}{!}{%
\begin{tikzpicture}[>=Stealth, auto,
    region/.style={circle, draw, font=\bfseries, minimum size=1.0cm, inner sep=0pt},
    main region/.style={circle, draw, fill=gray!20, font=\bfseries, minimum size=1.0cm, inner sep=0pt},
    label style/.style={font=\small}]

    \node[font=\scriptsize\bfseries, align=center, text width=2.8cm] at (-0.15,3.15) {(A) Direct loop\\($m=1$)};
    \node[main region] (jA) at (0,0.7) {Region $j$};
    \path[->, thick] (jA) edge [loop left, looseness=8, in=140, out=220] node[left, label style] {$\psi_{jj}$} (jA);

    \node[font=\scriptsize\bfseries, align=center, text width=3.5cm] at (4.6,3.15) {(B) Multi-regional\\loop ($m \ge 2$)};
    \node[main region] (jB) at (3.2,0.7) {Region $j$};
    \node[region] (k) at (5.8,1.95) {Region $k$};
    \node[region] (h) at (5.8,-0.55) {Region $h$};
    \path[->, thick] (jB) edge [bend left=10] node[above, label style] {$\psi_{kj}$} (k);
    \path[->, thick, dashed] (k) edge node[right, label style, yshift=2pt] {$\psi_{hk}$} (h);
    \path[->, thick] (h) edge [bend left=10] node[below, label style] {$\psi_{jh}$} (jB);

    \node[font=\scriptsize\bfseries, align=center, text width=3.2cm] at (10.0,3.15) {(C) Sensitivity\\duality};
    \node[draw, rounded corners, fill=gray!5, inner sep=8pt, align=center, text width=4.5cm] (eq) at (10.0,0.72) {
        \Large $\displaystyle \frac{\partial \rho(\boldsymbol\Psi(r))}{\partial \psi_{jj}(r)} = \frac{1}{E_j[m]}$ \\[8pt]
        \small Long Recurrence Time ($E_j[m] \gg 1$)\\
        $\Downarrow$\\
        Vanishing Sensitivity
    };

\end{tikzpicture}%
}
\caption{\textbf{Schematic representation of spatial recurrence loops and sensitivity duality.}
(A) Self-recurrence in region $j$ occurring within a single generation ($m=1$).
(B) Multiregional genealogical paths where a line of descent leaves region $j$ and returns after $m$ generations via other regions ($k,h$).
(C) The exact mathematical inverse relationship between the dominant-eigenvalue sensitivity $\partial \rho(\boldsymbol\Psi(r)) / \partial \psi_{jj}(r)$ and the mean recurrence generation $E_j[m]$.}
\Description{A schematic diagram contrasts a direct self-loop, a multiregional return loop through other regions, and the inverse relation between diagonal sensitivity and mean recurrence generation.}
\label{fig:conceptual_diagram}
\end{figure}

The quantity $\kappa_j$ also gives the sensitivity of the dominant eigenvalue $\rho(\boldsymbol\Psi(r))=1$, where $\boldsymbol\Psi(r):=(\psi_{ij}(r))$, to the diagonal element $\psi_{jj}(r)$:
\begin{equation}
\frac{\partial}{\partial\psi_{jj}(r)}\rho(\boldsymbol\Psi(r))=\kappa_j.
\end{equation}
At replacement level $r=0$, $\boldsymbol\Psi(0)$ is the next-generation matrix and its net reproduction rate is $\rho(\boldsymbol\Psi(0))$. Here the derivative varies $\psi_{jj}(0)$ while holding all other next-generation entries fixed; it is distinct from the sensitivity of the five-year population multiplier to age-specific survival or fertility in Eqs.~\eqref{sensK} and \eqref{sensF}. Greater diagonal sensitivity is equivalent to shorter mean recurrence and a larger normalized product of stable-birth share and reproductive value. It does not imply a larger stable-birth share by itself when reproductive values differ across prefectures.
Figure~\ref{fig:conceptual_diagram} provides a schematic guide to the notation for direct self-recurrence, multiregional recurrence loops, and the inverse relationship between diagonal sensitivity and mean recurrence generation.

\FloatBarrier

\subsection*{Population Replenishment Through Exogenous Inflows}

As a separate application of the same multiregional operator, we also examine population replenishment through repeated positive exogenous inflows. The purpose is to calculate, for a region experiencing marked decline, the inflow required to maintain its current reproductive-age population together with the cohort entering reproductive ages during the next five years, and then to evaluate the stationary population formed inside and outside the target prefecture when that inflow is repeated through fertility, survival, and internal migration. The procedure applies to any target prefecture. Let $h\in\{1,\ldots,M\}$ denote the target, and let $\mathbf M$ be the multiregional Leslie matrix constructed from observed 2020 fertility, survival, and internal migration. We permit only positive female inflows into prefecture $h$. These exogenous additions do not specify the entrants' regions of origin or nationality. Define $\mathbf P_{\mathrm{rep}}^{(h)}$ as the target vector retaining only the 2020 female population of prefecture $h$ from ages 10--14 through 45--49 and setting all other ages and prefectures to zero. Ages 10--14 are included because this cohort enters reproductive age during the next five years. The one-period residual required to maintain the target is
\begin{equation}
\mathbf q_{\mathrm{raw}}^{(h)}:=(\mathbf I-\mathbf M)\mathbf P_{\mathrm{rep}}^{(h)}.
\end{equation}
We exclude negative inflows, which would represent exogenous population removal, and retain only positive components:
\begin{equation}\label{immigrationq}
\mathbf q^{(h)}:=\max\{\mathbf q_{\mathrm{raw}}^{(h)},\mathbf0\},
\end{equation}
where the maximum is taken componentwise. The ages and sizes of positive inflows depend on the population and migration structure of $h$. If the same age-specific inflow is repeated every five years, dynamics follow
\begin{equation}
\mathbf n(t+1)=\mathbf M\mathbf n(t)+\mathbf q^{(h)}.
\end{equation}
Because the observed-rate matrix satisfies $\rho(\mathbf M)<1$, the stationary population is
\begin{equation}\label{immigrationstationary}
\lim_{t\to\infty}\mathbf n(t)=(\mathbf I-\mathbf M)^{-1}\mathbf q^{(h)}.
\end{equation}
This linear scenario holds the 2020 operator fixed and neither optimizes a policy nor estimates a causal effect. Truncating negative residuals means that the target vector itself need not be a stationary solution. The scenario assigns the positive part of the residual that maintains the target-age population for one period as a repeated exogenous inflow and displays the stationary population subsequently formed through births, survival, and internal migration.

\section*{Results}

\subsection*{Macro-Level Differences Between Classical and Multiregional Stable Population Models}

Table~\ref{tab:macro-comparison} compares basic demographic indicators from a classical Leslie matrix based on national averages with those from the multiregional Leslie matrix. We construct the single-region national model from final vital statistics \citep{MHLWVitalStatistics} and life tables \citep{MHLWLifeTable2010,MHLWLifeTable2015,MHLWLifeTable2020}, then divide fertility by the net reproduction rate so that national net reproduction equals one and the intrinsic growth rate is zero. For both models, we calculate the stationary female population by applying the stable-population projection in Eq.~\eqref{total} to the female population in each census year. Because reproductive value is zero at ages 50 and older, only women ages 0--49 contribute to the projection coefficient $\mathbf v\mathbf n_0$. The net reproduction rate in Table~\ref{tab:macro-comparison} is the multiregional net reproduction rate before normalization. Mean generation time $T$ and life expectancy $e_0$ are multiplied by five and reported in years.

Under replacement fertility, the stationary female population declines monotonically across census years in the national-average model but rises in 2020 in the multiregional model. A national average combines women living in regions with different fertility, survival, and migration schedules. It does not retain which regional life courses lead to births elsewhere or where those daughters subsequently reproduce. The multiregional calculation preserves these connections and the regional composition of the initial population. For example, although the unnormalized net reproduction rate is lowest in 2010, the multiregional stationary population is smallest in 2015. Thus, the national net reproduction rate alone does not order the stationary population totals. Table~\ref{tab:macro-comparison} establishes the aggregate contrast; the following prefectural profiles and recurrence paths identify the regional structure behind it.

\begin{table}[!htbp]
    \centering
    \caption{Observed and replacement-level stationary female population, Japan, 2010--2020}
    \label{tab:macro-comparison}
    \small
    \begin{tabular}{lrrrrrr}
        \toprule
        Year &
        \shortstack{Observed\\Female\\Population} &
        \shortstack{Classical\\Replacement\\Population} &
        \shortstack{Multiregional\\Replacement\\Population} &
        \shortstack{Net\\Reproduction\\Rate} &
        \shortstack{Mean\\Generation\\Time} &
        \shortstack{Life\\Expectancy\\at Birth} \\
        & \multicolumn{3}{c}{Millions} & & \multicolumn{2}{c}{Years} \\
        \midrule
        2010 & 65.7 & 53.8 & 61.8 & 0.647 & 33.1 & 86.1 \\
        2015 & 65.3 & 51.2 & 60.3 & 0.681 & 33.6 & 86.4 \\
        2020 & 64.8 & 48.9 & 62.6 & 0.675 & 34.2 & 86.6 \\
        \bottomrule
    \end{tabular}
    \begin{flushleft}
    \footnotesize
    Notes: Population counts are shown in millions of women. Both replacement populations are obtained by applying the stable-population projection in Eq.~\eqref{total} to the census-year female population. The classical benchmark is the single-region national Leslie model; the multiregional value is summed across all ages and 47 prefectures. Reproductive value is zero at ages 50 and older, so only women ages 0--49 contribute to the projection coefficient. The net reproduction rate is the dominant eigenvalue of the multiregional next-generation matrix before replacement-level normalization. Mean generation time and life expectancy at birth are calculated from the multiregional model and multiplied by five to express the five-year age-class results in years.
    \end{flushleft}
\end{table}

\FloatBarrier

\subsection*{Regional Characteristics and Disaster-Related Migration}

\begin{figure}[p]
    \centering
    \includegraphics[width=\textwidth,height=0.68\textheight,keepaspectratio]{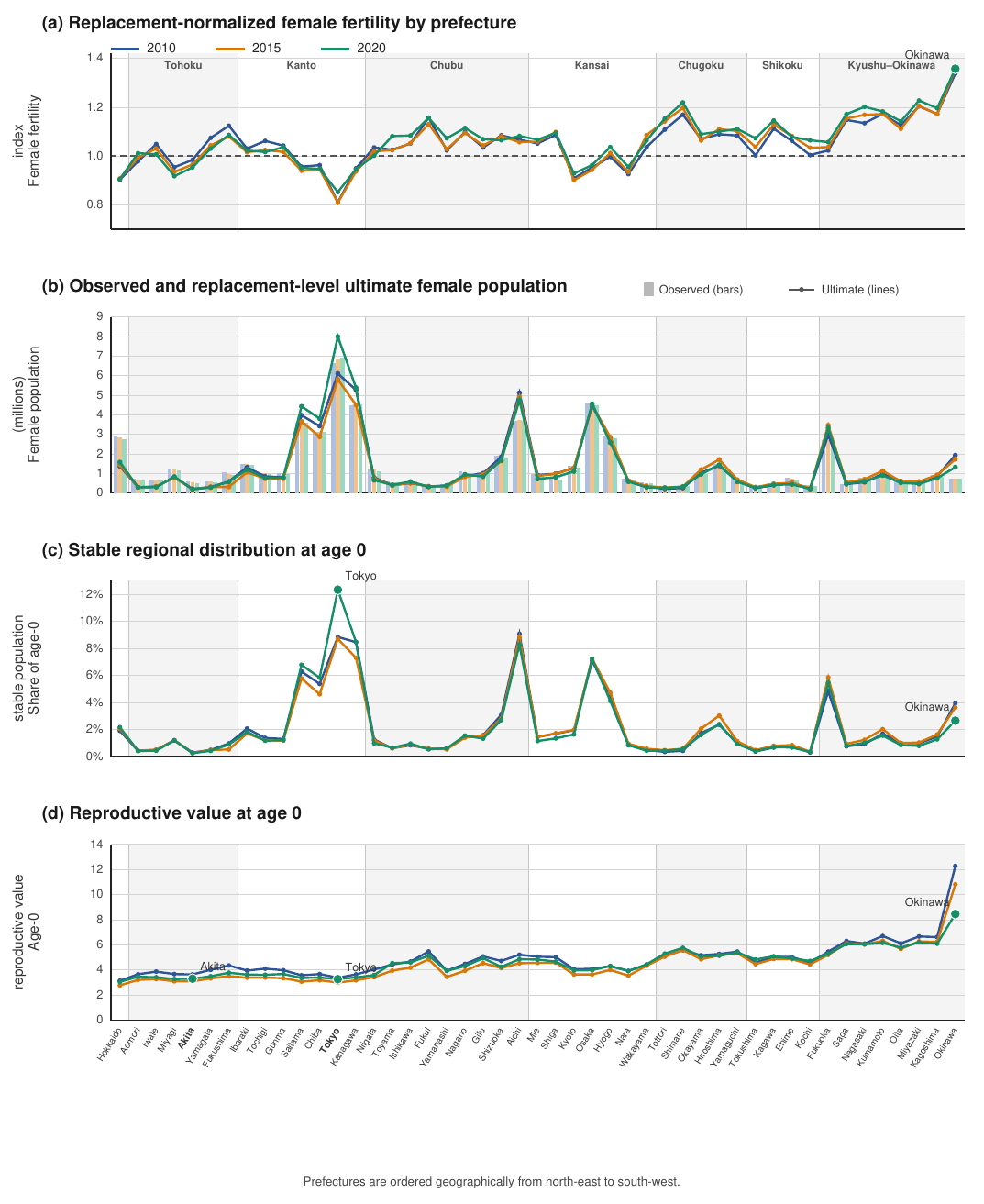}
    \caption{Prefectural profiles under replacement fertility. Panel a shows the prefectural female fertility index $I_i^{\mathrm F}$ after normalization of the national net reproduction rate to one (Eq.~\eqref{rffi}); panel b, observed female population (bars) and stationary female population under replacement fertility (lines); panel c, prefectural shares of age-zero population in the stationary population; and panel d, reproductive value at age zero. The reference line of one in panel a is not the replacement level of the conventional TFR. Prefectures are ordered geographically from northeast to southwest; colors indicate the reference year (2010, 2015, or 2020).}
    \Description{Four panels compare normalized female fertility, observed and stationary female populations, newborn shares, and reproductive values across Japan's 47 prefectures in 2010, 2015, and 2020.}
    \label{fig:selfreplace}
\end{figure}

In panel a of Figure~\ref{fig:selfreplace}, the normalized female fertility index $I_i^{\mathrm F}$ generally rises from northeastern to southwestern Japan. Okinawa has the highest value in all three years and lies well above the reference line of one. This does not mean that its conventional TFR exceeds one; it means that female fertility in Okinawa remains relatively high after applying the common multiplier that makes the national net reproduction rate equal one. This difference helps raise Okinawa's stationary population above its observed population under replacement fertility.

Panel b shows that Saitama, Chiba, Tokyo, and Kanagawa constitute the Tokyo metropolitan region. Except for Tokyo itself, these prefectures contain extensive commuter suburbs adjoining Tokyo's 23 central wards. Despite low fertility, their populations would increase substantially relative to observed levels if replacement fertility were reached under the 2020 demographic structure. Aichi, which contains Nagoya in central Japan, and Fukuoka, a major urban center in Kyushu, exhibit similar growth. By contrast, the northern prefectures from Hokkaido through Fukushima fall well below their observed populations. Akita combines low fertility with migration pathways that imply a substantially smaller stationary population even under the national replacement benchmark.

Fukushima experienced the Great East Japan Earthquake and the associated nuclear accident between the 2010 and 2015 censuses, followed by large-scale evacuation and outmigration. Against this documented migration disruption, its replacement-level stationary population calculated from the fixed 2015 rates is substantially below the corresponding 2010 value. The fixed 2020 rates imply a stationary level close to that calculated from 2010 rates. These are three separate period-model comparisons, not the trajectory of Fukushima's observed population or a simulation of its recovery. The comparison of recurrence below examines how the reproductive return structure also differs across those operators.

Panel c displays the prefectural distribution of the stable population at age zero. This distribution represents not only regional shares of newborns but also the contribution of women of any origin to descendants in each region. Akita has the smallest value in every census year, whereas the metropolitan prefectures surrounding Tokyo have the largest. Because female fertility in Akita exceeds that in Tokyo in panel a, fertility alone cannot account for the difference. The following analyses decompose it into one-generation birth contributions, intergenerational recurrence, and life-course migration pathways.

Panel d displays prefectural reproductive value at age zero under replacement fertility. Reproductive value generally rises toward southwestern Japan, and Okinawa is highest in every year. Under the genealogical interpretation, if the observed fertility, survival, and migration structure were repeated, lineages originating in Okinawa would gain relative influence as ancestors of Japan's future population.

\subsection*{One-Generation Regional Birth Contributions}

For fixed origin $j$, $\psi_{ij}(0)$ is the one-generation contribution of women originating in $j$ to births in region $i$. We define
\begin{equation}
O_j:=1-\frac{\psi_{jj}(0)}{\sum_{i=1}^{M}\psi_{ij}(0)}
\end{equation}
as the share of one-generation birth contributions occurring outside the prefecture of origin. Panel b of Figure~\ref{fig:migration-genealogy} shows the highest share for Kyoto; in Akita, approximately 60\% of the contribution is to births elsewhere. This share concerns daughters' birthplaces, not the fraction of women who migrate. The within-prefecture contribution $\psi_{jj}(0)$ shown in Figure~\ref{fig:self-prefecture-tfr}, panel a, is the first-return weight at $m=1$ under replacement, where $r=0$. Its broadly similar, low values for Tokyo, Akita, and Kyoto make the subsequent paths decisive for explaining their different mean recurrence generations. The outside-born daughters initiate the later steps of those paths through their own survival, movement, and reproduction.

\subsection*{Generational Migration Genealogies and Metropolitan Recurrence Loops}

\begin{figure}[p]
    \centering
    \includegraphics[width=\textwidth,height=0.78\textheight,keepaspectratio]{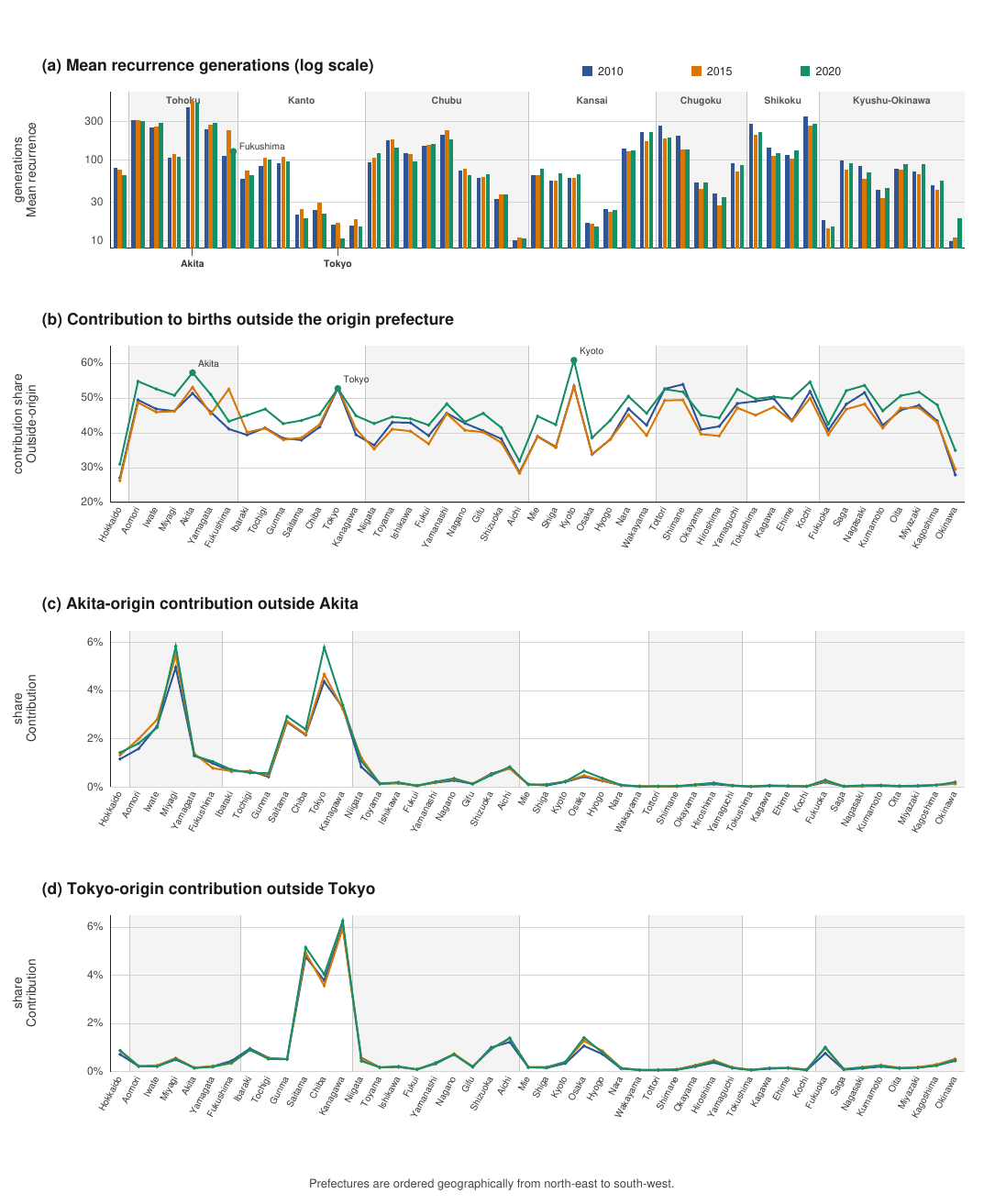}
    \caption{Intergenerational migration genealogies and metropolitan recurrence loops. Panel a shows the mean recurrence generation (log scale); panel b, the share of one-generation birth contributions outside the prefecture of origin, $O_j$; panel c, destination-specific birth contributions outside Akita by women originating in Akita; and panel d, destination-specific birth contributions outside Tokyo by women originating in Tokyo. Colors indicate the reference year (2010, 2015, or 2020).}
    \Description{Four panels show prefectural recurrence generations, outside-origin birth contributions, and the destinations of contributions by women originating in Akita and Tokyo.}
    \label{fig:migration-genealogy}
\end{figure}

Panel a of Figure~\ref{fig:migration-genealogy} displays the mean recurrence generation $E_j[m]$ on a logarithmic scale. Based on 2020 period rates, it is approximately 10 generations in Tokyo and approximately 500 in Akita. Mean recurrence is the expected first-return generation under repeated application of the period next-generation operator. Akita's large value identifies weak closed reproductive pathways back to Akita at the observation date. Like stable population and generation time, this is a structural period quantity, not a forecast horizon. Fukushima's mean recurrence is much higher under the fixed 2015 operator than under the 2010 operator and lower again under the 2020 operator. Together with Figure~\ref{fig:selfreplace}, panel b, this comparison identifies changes in the period reproductive structure, rather than a predicted waiting time for postdisaster recovery.

Panel c of Figure~\ref{fig:migration-genealogy} shows the destinations of Akita-origin contributions outside Akita. Contributions to other prefectures in Tohoku, the northeastern region of Japan, are comparatively high, but those to the Tokyo metropolitan region equal or exceed them. A daughter born in Tokyo from an Akita-origin mother then contributes to the next reproductive generation through the Tokyo-origin schedule. Panel d shows the outside-Tokyo destinations in that schedule. Reading the two panels in sequence therefore connects a mother's life course to the subsequent births of her descendants.

The Tokyo-origin contributions outside Tokyo are concentrated in Saitama, Chiba, and Kanagawa, with smaller contributions to Akita (Figure~\ref{fig:migration-genealogy}, panel d). Conversely, the contributions to births in Tokyo from these neighboring prefectures are prominent in Figure~\ref{fig:self-prefecture-tfr}, panel b-1. These paired directions support short reproductive loops through the metropolitan region. Akita-origin daughters enter this same network, but the contributions back to Akita are small (panel b-3). Similar direct self-contributions can therefore coexist with very different weights on later first-return generations. The displayed one-generation profiles explain the direction of this contrast; $E_j[m]$ in Eq.~\eqref{em} combines all first-return paths, including those through prefectures beyond these examples.

\FloatBarrier

\subsection*{Life-Course Migration Profiles}

Having characterized intergenerational recurrence through the next-generation matrix $\boldsymbol\Psi(r)$, we now examine the life-course survivorship $K_{ij}(a\mid s)$ from which the one-generation reproductive contributions are constructed. Tokyo and Akita provide contrasting cases: under the 2020 period operator, Tokyo has short recurrence despite low fertility, whereas Akita has long recurrence and a much smaller replacement-level stationary population.

\begin{figure}[p]
    \centering
    \includegraphics[width=\textwidth,height=0.68\textheight,keepaspectratio]{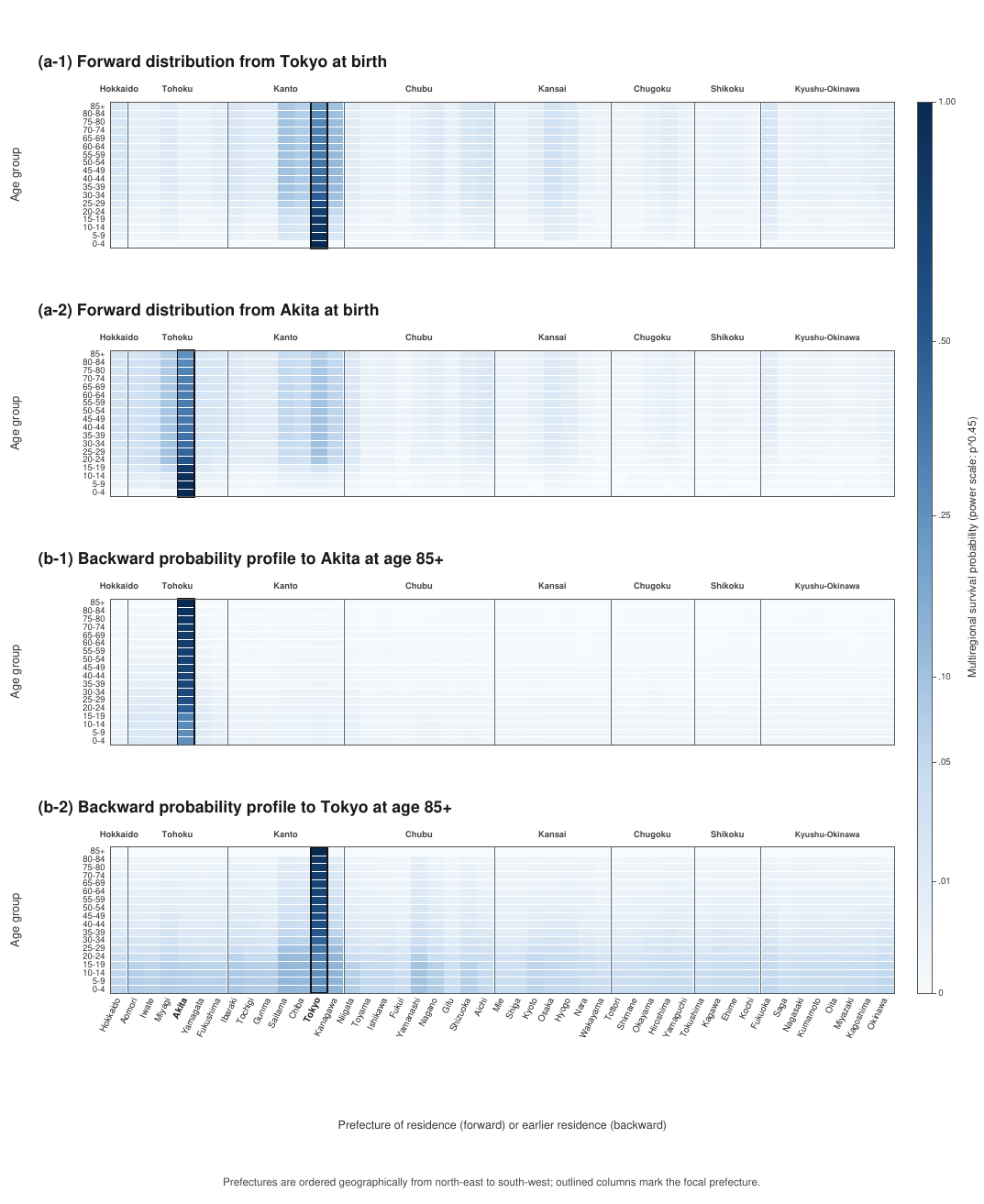}
    \caption{Life-course migration profiles from multiregional survivorship in 2020. Panels a-1 and a-2 show forward distributions for women originating in Tokyo and Akita, respectively. Panels b-1 and b-2 show backward profiles leading to residence at age 85 or older in Akita, $K_{5j}(85+\mid s)$, and Tokyo, $K_{13j}(85+\mid s)$, respectively. A backward profile is the multiregional probability that a woman residing in prefecture $j$ at age $s$ survives and resides in the target prefecture at age 85 or older. To reveal small regional differences, color intensity uses the nonlinear transformation $p^{0.45}$ of probability $p$.}
    \Description{Heat maps contrast forward migration from Tokyo and Akita with backward probabilities of reaching either prefecture at age 85 or older from earlier ages and residences.}
    \label{fig:life-course-migration}
\end{figure}

Panels a-1 and a-2 of Figure~\ref{fig:life-course-migration} are forward distributions of the multiregional survivorship in Eq.~\eqref{dk1}, evolved from $s=0$ over age $a$ for Tokyo and Akita. Most women originating in Tokyo either remain there or move within the metropolitan region; few move to nonmetropolitan cities. This is consistent with Tokyo-origin contributions to reproduction elsewhere. Women originating in Akita, by contrast, move not only within Tohoku but also to the Tokyo region and thus have more geographically varied residence trajectories. Across panel a, within-generation migration is closely connected to reproductive contribution by region.

Using Eq.~\eqref{backwardprofile}, panel b shows the probability of eventually residing in Akita or Tokyo at age 85 or older by age $s$ and residence $j$ at that age. In panel b-1, women residing in Akita have the highest probability of remaining there at age 85 or older, whereas the probabilities of reaching Akita from other Tohoku prefectures or Tokyo are comparatively small. In panel b-2, the probability of residing in Tokyo at age 85 or older is positive across a broad range of earlier residences. Survivorship-migration pathways to Tokyo therefore exist from many parts of Japan. These probabilities condition on earlier residence; they are not the composition of Tokyo's older population by origin.

In summary, the forward distribution for women originating in Akita contains multiple destinations in neighboring regions and the Tokyo metropolitan area, whereas the backward profile ending in Akita at age 85 or older is strongly concentrated among women who previously lived there. The probability of ending in Tokyo at age 85 or older is geographically broad across previous residences. This contrast demonstrates regional differences in lifetime migration pathways. Birth contributions additionally weight such pathways by fertility at reproductive ages; residence at age 85 or older is not itself a reproductive return.

\FloatBarrier

\subsection*{Contrasting Regional Reproduction Regimes}

\begin{figure}[p]
    \centering
    \includegraphics[width=\textwidth,height=0.80\textheight,keepaspectratio]{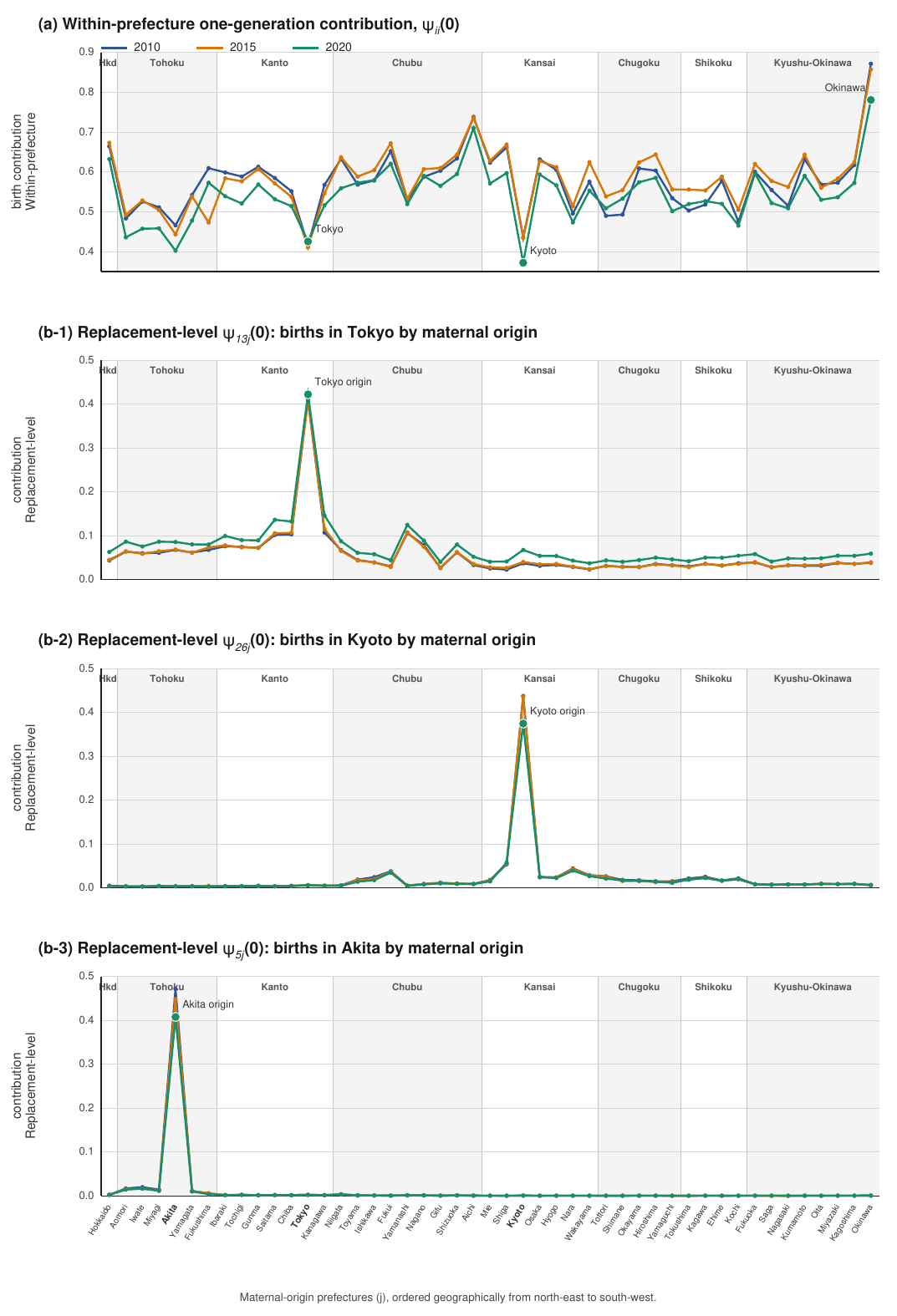}
    \caption{One-generation interregional birth contributions under replacement fertility. Panel a shows within-origin-prefecture contributions $\psi_{ii}(0)$. Panels b-1, b-2, and b-3 show contributions by maternal prefecture of origin $j$ to births in Tokyo, $\psi_{13j}(0)$; Kyoto, $\psi_{26j}(0)$; and Akita, $\psi_{5j}(0)$, respectively. Prefectures are ordered geographically from Hokkaido to Okinawa; colors indicate the reference year (2010, 2015, or 2020).}
    \Description{Four panels compare within-prefecture birth contributions and origin-specific contributions to births in Tokyo, Kyoto, and Akita for three census years.}
    \label{fig:self-prefecture-tfr}
\end{figure}

Tokyo and Akita primarily illustrate differences in return pathways, whereas Hokkaido and Okinawa show how local fertility and within-region reproduction can generate different stationary outcomes even without strong metropolitan inflows. The preceding results show how multiregional survivorship over individual life courses and the resulting intergenerational path weights in Eq.~\eqref{pi} form stationary populations through migration genealogies. In addition to Akita and Tokyo, panel a of Figure~\ref{fig:self-prefecture-tfr} highlights Hokkaido and Okinawa, whose stationary populations diverge through different mechanisms. Both have high within-prefecture one-generation birth contributions $\psi_{ii}(0)$. Hokkaido is a comparatively populous northern region containing Sapporo, a government-designated major city, whereas Okinawa's total population is smaller than Sapporo's alone. Panel b of Figure~\ref{fig:selfreplace} shows that Hokkaido's stationary female population falls to about half its observed population under replacement fertility, while Okinawa's nearly doubles. Okinawa also consistently has a high female fertility index in panel a of Figure~\ref{fig:selfreplace}. To assess whether local fertility can sustain growth without interregional inflows, we compare Okinawa with a closed single-region model. When interregional migration is removed after national normalization and a closed single-region model is constructed using only Okinawa's fertility and survival, its net reproduction rate is approximately 1.35, above replacement. Okinawa has a structure in which local fertility supports long-run population growth.

Kyoto sharpens the Tokyo--Akita comparison. It has the largest outside-origin share in Figure~\ref{fig:migration-genealogy}, panel b, and a low direct self-contribution in Figure~\ref{fig:self-prefecture-tfr}, panel a, yet its mean recurrence is shorter than Akita's. Contributions to births in Kyoto come from other prefectures, especially central and western Japan (panel b-2), whereas the corresponding profile for Akita is strongly concentrated on Akita-origin women (panel b-3). Tokyo has broader contributions from other origins (panel b-1). The amount of reproduction occurring outside the origin prefecture consequently does not by itself determine recurrence: the subsequent connections of those birth regions also matter. The identity in Eq.~\eqref{em} relates the resulting mean to stable births and reproductive value jointly, not to the stationary population share alone.

\FloatBarrier

\subsection*{Demographic Implications}

Figure~\ref{fig:age-distribution-reproductive-value} returns from the regional pathway comparisons to the full age-by-prefecture system. It compares stable distributions and reproductive values under the replacement-normalized and observed 2020 operators and then examines how age-specific survival--migration and fertility entries contribute locally to the population growth multiplier. In reality, every prefecture remained below replacement in 2020. In the replacement-level stable distribution (Figure~\ref{fig:age-distribution-reproductive-value}, panel a-1), Tokyo has the largest population around the early 20s because of inflows for education and employment. Because nationwide low fertility is removed, age distributions are otherwise relatively flat apart from such regional features. Under the observed $r<0$ dynamics that retain low fertility (Figure~\ref{fig:age-distribution-reproductive-value}, panel a-2), the center of population shifts toward older ages in every prefecture.

The reproductive-value profiles are similar (Figure~\ref{fig:age-distribution-reproductive-value}, panels b-1 and b-2), although observed values are higher in southwestern prefectures than under replacement. Panel c uses Eq.~\eqref{sensK} and \eqref{sensF} to identify the prefecture and age associated with the maximum absolute sensitivity of the five-year growth multiplier $\lambda$. In the 2020 data, the maximum sensitivity to $k_{ij}(a)$ occurs for movement from Tokyo to Okinawa, both under observed and replacement-level fertility. Migration sensitivity is larger through the 20s, whereas sensitivity to diagonal fertility $f_{ii}(a)$ becomes larger from the 30s onward. This age crossover identifies different age-specific contributions to the population growth multiplier. Because matrix entries are perturbed independently and no behavioral response or feasible policy constraint is modeled, it does not identify the causal effectiveness of migration or fertility interventions. The result extends the analysis through 2015 in \citet{Oizumi2022PLOS}.

\subsection*{Application: Positive Exogenous Inflows}

Finally, we consider replenishment by positive exogenous inflows under the fixed 2020 demographic operator. This is a linear replenishment scenario, not an optimal policy or a causal intervention estimate. Because Akita exhibits the greatest decline in Figure~\ref{fig:selfreplace}, panel b, we select it empirically and set $h=\mathrm{Akita}$ in the general procedure. Write the resulting vector as $\mathbf q^{\mathrm{Akita}}:=\left.\mathbf q^{(h)}\right|_{h=\mathrm{Akita}}$.

\begin{figure}[p]
    \centering
    \includegraphics[width=\textwidth,height=0.42\textheight,keepaspectratio]{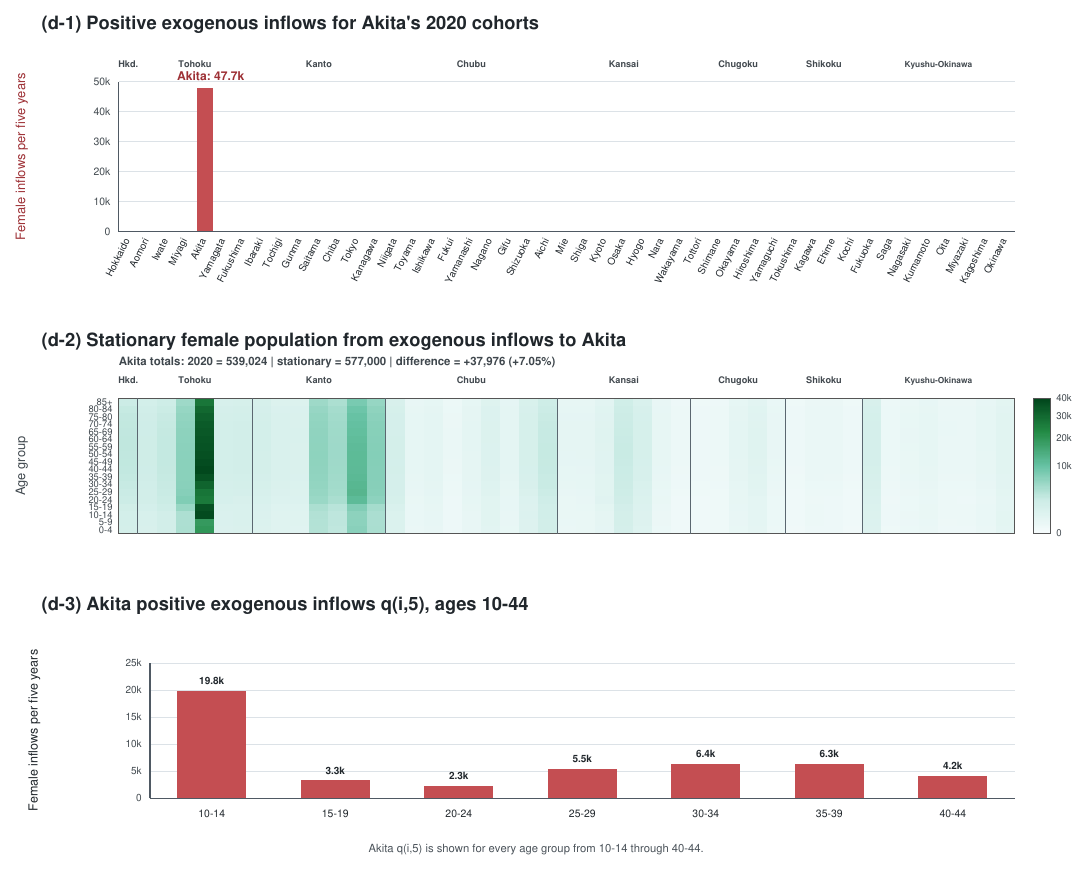}
    \caption{Stable age distributions, reproductive values, sensitivities, and the stationary population generated by positive exogenous female inflows to Akita, by prefecture and age class in 2020. Panels a-1 and a-2 show the replacement-level stable age distribution and the observed-rate stable distribution $w(a,i)$; panels b-1 and b-2, reproductive value at replacement level and observed reproductive value $v(a,j)$; and panel c, the maximum absolute sensitivity of the five-year population multiplier $\lambda$ to multiregional survival and the absolute sensitivity to diagonal fertility in Tokyo and Aichi. Sensitivities vary each matrix element independently without imposing the column-sum constraint on migration probabilities. Panel d-1 shows the five-year total positive female inflow at ages 10--44 obtained from Eq.~\eqref{immigrationq} using a target vector that retains only Akita's 2020 female population at ages 10--49. Panel d-2 shows the stationary female population over all ages and prefectures from Eq.~\eqref{immigrationstationary}; labels report Akita's 2020 population (539,024), stationary population (577,000), and difference ($+37,976$; $+7.05\%$). Panel d-3 shows positive inflow $q^{\mathrm{Akita}}(a,5)$ by age from 10--14 through 40--44. Inflow is assigned only to Akita, but births and multiregional survivorship distribute the stationary population elsewhere. Reproductive value is zero at ages 50 and older. Rows in panels a, b, and d-2 are age classes and columns are prefectures ordered from Hokkaido to Okinawa. Colors in panels a-1, a-2, and d-2 transform each value relative to its panel maximum as $x^{0.5}$; panels b-1 and b-2 use a linear scale. Survival sensitivity is shown for ages 0--49 and fertility sensitivity for ages 15--49.}
    \Description{Heat maps and line plots compare stable age structure, reproductive value, age-specific sensitivities, and the population generated by repeated positive female inflows to Akita.}
    \label{fig:age-distribution-reproductive-value}
\end{figure}

The required positive female inflows from Eq.~\eqref{immigrationq} occur from ages 10--14 through 40--44 and total approximately 47,741 over five years (Figure~\ref{fig:age-distribution-reproductive-value}, panel d-1); inflow at ages 45--49 is zero (panel d-3).

Assuming a 1:1 sex ratio and accepting men with the same age structure gives a total of approximately 95,483 people.

Under repeated inflows every five years, the full multiregional system produces a stationary female population of approximately 1,753,049 across all prefectures.

Within Akita, the stationary female population is approximately 577,000. This is 37,976 (7.05\%) above its 2020 population of 539,024. Of the stationary population formed from exogenous entrants and their descendants, 67.1\% is located outside Akita through repeated internal migration pathways (Figure~\ref{fig:age-distribution-reproductive-value}, panel d-2). Young inflows that include ages 10--14 shape long-run population both within and outside Akita through subsequent fertility and migration.

\clearpage
\begin{figure}[p]
    \centering
    \includegraphics[width=\textwidth,height=0.76\textheight,keepaspectratio]{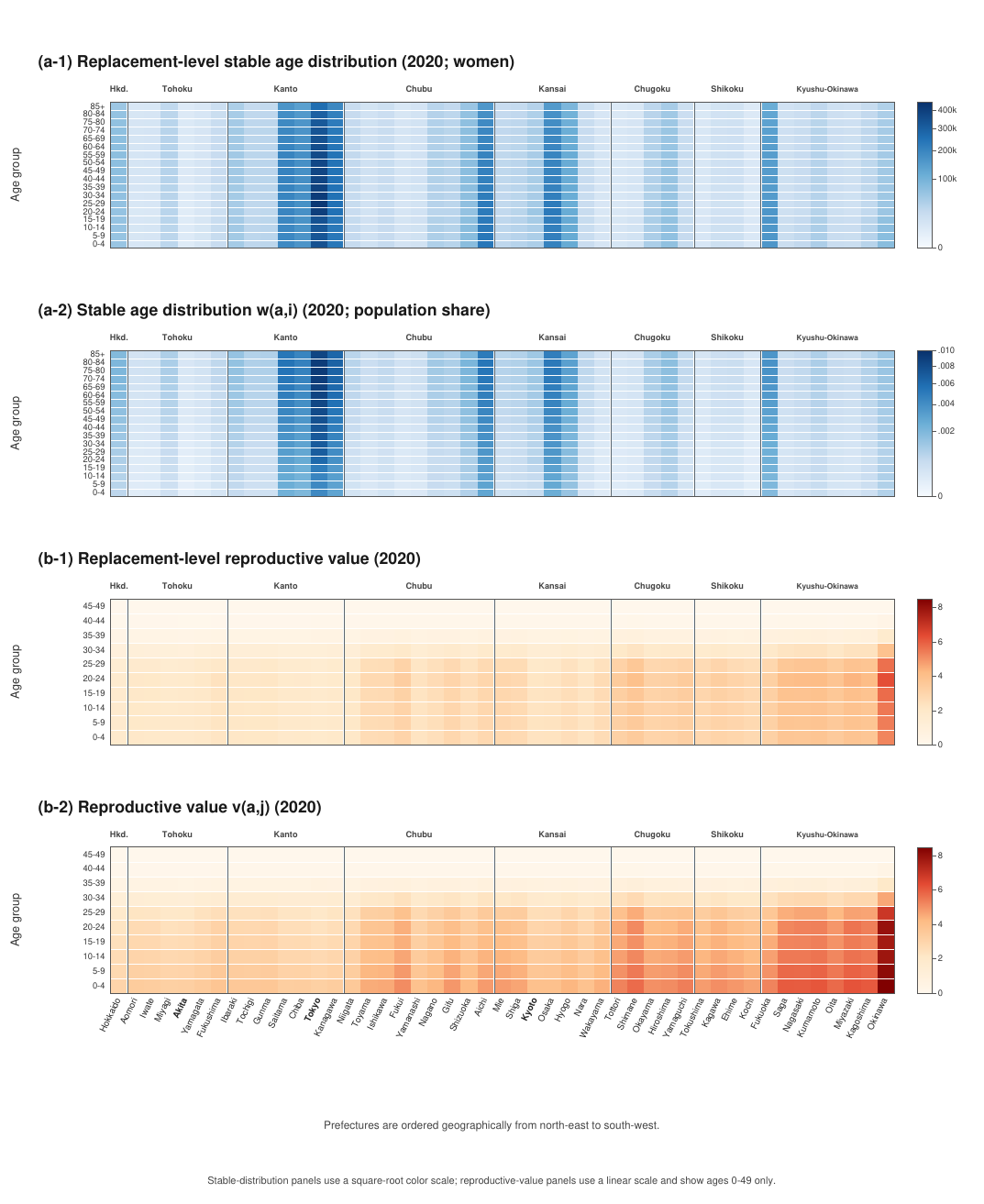}
    \par\smallskip{\small\itshape Figure~\ref{fig:age-distribution-reproductive-value} continues on the next page.}
\end{figure}

\clearpage
\begin{figure}[p]
    \centering
    \includegraphics[width=\textwidth,height=0.38\textheight,keepaspectratio]{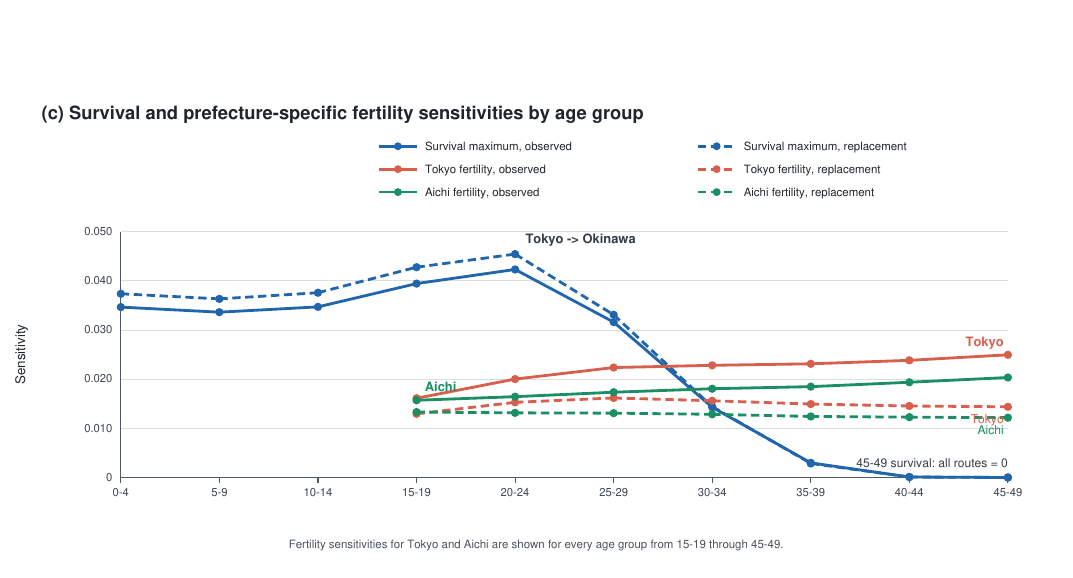}
    \par\smallskip{\small\itshape Figure~\ref{fig:age-distribution-reproductive-value} continues on the next page.}
\end{figure}

\clearpage

\FloatBarrier

\section*{Discussion}

A migration genealogy is not a simple extrapolation of period net migration. It is a recurrent structure in which age-specific survival and migration paths are converted through births into the birth regions of the next generation. This genealogical interpretation and the series for the stable age distribution and reproductive value follow \citet{Oizumi2022PLOS}; our recurrence identities establish demographic relationships within those series. The first-return distribution and the identity linking stable births to reproductive value connect these pathways to mean recurrence. At replacement, mean recurrence equals the reciprocal sensitivity of net reproduction to within-region reproduction. This demographic identity links regional reproductive loops to the response of population reproduction.

Classical multiregional theory anticipates uneven regional replacement; the present decomposition identifies the first-return genealogies underlying that unevenness within the specified period operator and links their mean exactly to stable births, reproductive value, and the sensitivity of net reproduction at replacement. The Japanese application demonstrates these mechanisms through Akita, Tokyo, and Kyoto. Their within-region one-generation contributions are broadly similar and low, but descendant destinations and subsequent reproductive pathways differ, producing different first-return distributions and mean recurrence generations. The inherited path series resolve how these pathways compose stable births and reproductive value; the mean-recurrence identity connects their normalized product to the strength of closed reproductive pathways. These comparisons thus expose the demographic mechanisms generating regional differences under the period operator, beyond what net migration or one-generation contributions alone reveal.

The contrast between Tokyo and Akita illustrates the mechanism. Tokyo's low fertility is partly offset by inflows of people at reproductive ages and short recurrence pathways concentrated within the metropolitan region. Movement from Akita is directed toward neighboring prefectures and the Tokyo region, but genealogical pathways back from those destinations are weak. Akita has a small relative population under replacement even though its fertility exceeds Tokyo's. The life-course profiles show the survival and movement underlying these birth contributions: Akita-origin cohorts have destinations in Tohoku and the Tokyo region, whereas pathways reaching Akita from other prefectures are weak. Kyoto further shows that a large outside-origin birth share need not imply recurrence as long as Akita's when subsequent reproductive connections differ. The relevant distinction is not simply urban versus rural but the reproductive pathways into which a region is embedded. As Hokkaido and Okinawa demonstrate, sufficiently high local fertility can support population even without strong inflows, producing multiple demographic regimes from different combinations of fertility and migration.

Comparisons across 2010, 2015, and 2020 show that this structure is not fixed. In Fukushima, the period rates spanning the disaster and nuclear accident imply a sharply smaller stationary population and longer recurrence in 2015, with both measures moving back toward their 2010 values under 2020 rates. This is consistent with the documented disruption of migration systems, although the comparison does not isolate migration from the simultaneous changes in fertility and survival. Tokyo's mean recurrence and stationary population based on 2020 rates likewise differ from those based on 2010 and 2015. Our stationary populations are not point forecasts of the distant future but comparative indicators of the long-run structure implied by fixing the demographic processes observed at each date. Akita's exceptionally long mean recurrence does not predict unchanged behavior for hundreds of generations; it measures the scarcity of return pathways in the observed network.

These results support considering fertility, migration, and regional conditions together: a birth in one prefecture can contribute to reproduction elsewhere when daughters move before childbearing, and inflows can have consequences beyond the receiving prefecture. The relevant questions for regional population planning include where entrants later reproduce and how their descendants connect to other regions. The model describes these demographic connections without estimating how employment, housing, or family support would change them. The recurrence--sensitivity relation describes this dependence locally, while the Akita inflow scenario in Figure~\ref{fig:age-distribution-reproductive-value} describes a separate linear application under fixed 2020 rates. Repeated positive female inflows at ages 10--44 yield a stationary female population 7.05\% above Akita's 2020 level, with 67.1\% of the resulting stationary population located outside Akita. The same multiregional pathways that define recurrence also distribute the descendants of exogenous inflows beyond the target region. These quantities describe propagation through the specified operator. They do not establish optimal inflow levels, migrant retention, or the effectiveness of specific policies; such questions require behavioral evidence and feasible joint changes in demographic rates.

Several limitations qualify the analysis. First, we hold period fertility, survival, and migration rates from each census year constant indefinitely and do not endogenize economic cycles, policy change, postdisaster recovery, or temporary migration changes around 2020. Second, census residence five years earlier does not capture multiple moves within the interval, and migration histories are unavailable for decedents and international outmigrants. Age 85 and older is an aggregated terminal class, and subsequent retention within that class is not followed. Third, the female-dominant model consistently tracks female births and maternal migration but omits men as an explicit state. It cannot represent constraints from regional sex ratios and marriage-market imbalance, sex differences in migration timing and destination, tied migration, or couple- and household-level decisions. These processes may change the feasibility of births and direction of return migration, so stable populations and recurrence based on female rates alone may be biased upward or downward. Fourth, international migration is not an explicit regional state, preventing evaluation of inflow, settlement, and return among foreign-born populations. Fifth, prefectural aggregation obscures urban--rural differences within prefectures and functional ties across commuting zones. Future research should test migration genealogies and realistic policy scenarios using time-varying multiregional models, two-sex models coupling both populations and partner formation, multistate household models, open-population models with international migration, and finer geographic units.

\section*{Conclusion}

Decomposition into recurrence pathways elucidates the intergenerational reproductive mechanisms determining the stationary population distribution under national replacement. Starting from the existing migration-genealogy series, we establish a first-return probability structure for multiregional stable populations and derive its recurrence identities. The series connect individual life-course survival and migration to intergenerational migration genealogies repeated through births; these genealogical and adjoint contributions compose stable births and reproductive value. The first-return Euler--Lotka relation, mean recurrence, and reciprocal diagonal sensitivity connect these quantities in a single demographic structure. Classical multiregional theory anticipates that national replacement need not imply uniform regional replacement, but its aggregate description of the stable age distribution and reproductive value does not resolve the first-return genealogical mechanisms underlying those differences. Building on the existing series, we identify and quantify these mechanisms through the first-return probability law and the mean-recurrence identities. Under national replacement with migration, region-specific reproductive pathways yield different first-return distributions and compose different stable-birth and reproductive-value profiles, producing nonuniform regional replacement. The mean-recurrence identity links this return structure exactly to the normalized product of stable births and reproductive value and to reciprocal sensitivity of net reproduction to within-region reproduction at replacement.

Application to Japan's 47 prefectures in 2010, 2015, and 2020 measures regional reproductive pathways that fertility and net migration alone cannot characterize. Contrasts among Tokyo, Akita, and Kyoto trace the difference from similar within-region one-generation contributions to distinct destinations and subsequent birth pathways. Fukushima demonstrates that recurrence changes with the period operator. The demographic question is consequently not only where people move, but where their descendants are born and through which reproductive paths a lineage returns. Under appropriate irreducibility and convergence conditions, the same path-based framework can be applied to other multistate structured populations.

\section*{Acknowledgments}

The authors are grateful to Kumiko Oizumi, Shin Oizumi, Ko Oizumi, and Hiroko Oizumi for their support and encouragement. The authors also thank Hisashi Inaba for valuable advice, Youichi Enatsu for helpful discussions, and Dr. Yoshinori Kamijima for many valuable suggestions and advice.

OpenAI Codex was used for English translation from non-English-language text, English-language editing, and manuscript formatting and restructuring based on instructions supplied by the authors. The authors reviewed and revised all AI-assisted text and take full responsibility for the scientific content, mathematical derivations, data analysis, interpretation, conclusions, accuracy, arguments, and final wording of the manuscript. OpenAI Codex was not used to generate data, conduct the demographic analyses, select results, or create figures, and is not treated as an author.

\paragraph{Funding.}
This work was supported by Health, Labour and Welfare Sciences Research Grants from the Ministry of Health, Labour and Welfare of Japan (Grant Number JPMH26AA2009; project title: ``Population and household projections and social restructuring focusing on diversifying household structures''). Yuki Chino acknowledges support from the NSTC grant 111-2115-M-A49-009-MY3.

\clearpage
\section*{Data Availability}
Population and migration inputs come from the Population Census of Japan; fertility inputs come from Vital Statistics of Japan; survival inputs come from prefectural life tables. Appendix A and the references provide dataset identifiers and public source URLs.

\renewcommand{\refname}{References}

\clearpage
\section*{Appendix A: Input Data and Matrix Construction}

This appendix provides the inputs and transformations required to reproduce the model defined in the main text. The population consists of female residents of Japan in 47 prefectures and five-year census age classes $0$--$4,\ldots,80$--$84,85+$. Population vectors and stored matrices order age class in the outer dimension and prefecture in the inner dimension.

\subsection*{Input Statistics}

Base populations and interregional migration come from the tabulations of migration by sex and age in the 2010, 2015, and 2020 population censuses \citep{Census2010,Census2015,Census2020}. The 2010 and 2015 inputs use e-Stat classification IDs 000001048107 and 000001093875, and the 2020 input uses Table 1 of data set 000032168214. Sex ratio at birth, infant mortality, and births by maternal age come from final vital statistics \citep{MHLWVitalStatistics}. Female survival comes from the female prefectural life tables (tables 01002, $\ldots$, 47002) for each census year \citep{MHLWLifeTable2010,MHLWLifeTable2015,MHLWLifeTable2020}. We neither interpolate age classes nor prorate population counts.

\subsection*{Migration, Survival, and Fertility Components}

Let $P_y(x;j)$ be the female population of age $x$ in region $j$ at time $y$, and let $P_{y+5}(x;j\to i)$ be the number residing in region $i$ at time $y+5$ who resided in region $j$ five years earlier. For $i\neq j$, the five-year migration probability is
\begin{equation}
T_{ij}^{(y)}([a-5,a-1])=
\frac{\sum_{x=a}^{a+4}P_{y+5}(x;j\to i)}
{\sum_{x=a-5}^{a-1}P_y(x;j)},
\qquad
T_{jj}^{(y)}=1-\sum_{i\neq j}T_{ij}^{(y)}.
\end{equation}
For the oldest age class, we divide movers age 85 or older at the census by the population age 80 or older five years earlier and define the result as the transition probability into the aggregated terminal class, age 85 or older. This terminal treatment follows the published age categories. Survival-migration entries from age 85 or older to the next period are zero, so members of this class leave the model during the subsequent five years. We exclude records whose residence five years earlier was abroad, unknown, or not assignable to a domestic prefecture; the model is a closed-population approximation for the 47 prefectures. The probability describes the transition between the beginning and end of a five-year interval and does not observe multiple moves or intermediate paths within the interval.

Let $L_{x,j}^{(y)}$ be life-table survivors and $l_{x,j}^{(y)}=L_{x,j}^{(y)}/L_{0,j}^{(y)}$. Five-year survival and multiregional survival are
\begin{equation}
p_j^{(y)}(a)=\frac{l_{a+4,j}^{(y)}}{l_{a,j}^{(y)}},
\qquad
k_{ij}^{(y)}(a)=T_{ij}^{(y)}(a)p_j^{(y)}(a).
\end{equation}
We apply survival for the region of residence $j$ at the beginning of the interval.

Let $MSB_t$ be male births per 100 female births, $IFM_i(y)$ infant deaths per 1,000 births, and $B_{i,x}(t)$ births per 1,000 women age $x$. Following \citet{Oizumi2022PLOS}, fertility for five-year age class $[a,a+4]$ is
\begin{equation}
f_i^{(y)}(a)=
\frac{100}{100+\frac{1}{5}\sum_{t=y-4}^{y}MSB_t}
\left(1-\frac{IFM_i(y)}{1000}\right)
\frac{1}{1000}\sum_{k=0}^{4}B_{i,a+k}(y-k)l_{k,i}^{(y)}.
\end{equation}
Newborns are assumed to inherit their mother's region of residence, so $f_{ij}^{(y)}(a)=f_i^{(y)}(a)\delta_{ij}$.

\subsection*{Reproducibility and Interpretation}

For each year, we construct an $846\times846$ multiregional Leslie matrix, a $47\times47$ next-generation matrix, and an $18\times47$ initial female population array. Age class is the outer dimension and prefecture the inner dimension. Validation confirms $T_{ij}\geq0$, $\sum_iT_{ij}=1$, $0\leq p_j\leq1$, matrix nonnegativity, correct dimensions, and consistent index order. Each annual matrix represents a counterfactual that holds the demographic structure of the observed period fixed rather than a forecast. The 2015 matrix may incorporate the migration environment after the Great East Japan Earthquake, and the 2020 matrix may incorporate the early COVID-19 environment; differences across years include period-specific social disruption.

\clearpage
\section*{Appendix B: Eigenvector Series and Genealogies}

For the asymptotic statements, assume that the multiregional Leslie matrix is primitive, so its Perron root is simple and strictly dominates the moduli of the remaining eigenvalues. Irreducibility of the next-generation matrix is sufficient for the first-return series in Appendix B. The absolute-error form in the main text additionally assumes that the remaining eigenvalues of the Leslie matrix have modulus below one; without that stronger condition, Eq.~\eqref{sp1} remains the relevant Perron-normalized limit.

This appendix restates the eigenvector representation theorem in the main text and S1 File of \citet{Oizumi2022PLOS} in notation corresponding to the next-generation matrix $\boldsymbol\psi(r)$ used here. The right- and left-eigenvector series and their characteristic-equation relation below reproduce that existing representation, including paths that avoid the reference region at intermediate steps. We then use their first-return terms as a normalized probability law and establish Proposition B.1, the mean-recurrence identity, and its reciprocal-sensitivity consequence. Standard matrix and Perron--Frobenius results follow \citet{Meyer2000}.

\subsection*{Representation of the Right Eigenvector}

Let $A=(a_{ij})_{1\leq i,j\leq n}$ be an irreducible nonnegative matrix and let $\lambda>0$ be its Perron root. Set $B=A/\lambda$, so $\rho(B)=1$. A right eigenvector $w=(w(i))$ satisfies
\begin{equation}
w=Bw.
\end{equation}
Fix a reference state $\ell$ with $w(\ell)\neq0$. Let $w_{-\ell}$ be the vector without component $\ell$, let $Q_\ell$ be the principal submatrix of $B$ with row and column $\ell$ removed, and let $b_\ell$ be column $\ell$ of $B$ without its $\ell$th component. Then
\begin{equation}
w_{-\ell}=Q_\ell w_{-\ell}+w(\ell)b_\ell.
\end{equation}
Irreducibility implies $\rho(Q_\ell)<1$, and the Neumann series converges absolutely componentwise. Hence
\begin{equation}
w_{-\ell}=w(\ell)(I_\ell-Q_\ell)^{-1}b_\ell
=w(\ell)\sum_{m=0}^{\infty}Q_\ell^mb_\ell.
\end{equation}
Componentwise,
\begin{equation}
w(i)=w(\ell)\left(\lambda^{-1}a_{i\ell}
+\sum_{m=1}^{\infty}\sum_{j_1,\ldots,j_m\neq\ell}
\lambda^{-m-1}a_{ij_1}a_{j_1j_2}\cdots a_{j_m\ell}\right).
\end{equation}
This expression decomposes paths between $i$ and the reference state $\ell$ into those that do not visit $\ell$ at an intermediate step. Setting $A=\boldsymbol\psi(r)$ gives a migration-genealogy series: a birth lineage originating in region $j$ avoids returning to $j$, passes through other regions, and returns in a specified generation.

\subsection*{Recurrence as a Characteristic Equation}

The reference component satisfies
\begin{equation}
w(\ell)=\lambda^{-1}\sum_{j=1}^{n}a_{\ell j}w(j).
\end{equation}
Substituting the series representation gives
\begin{equation}
1=\lambda^{-1}a_{\ell\ell}
+\sum_{m=1}^{\infty}\sum_{j_1,\ldots,j_m\neq\ell}
\lambda^{-m-1}a_{\ell j_1}a_{j_1j_2}\cdots a_{j_m\ell}.
\end{equation}
As shown in the S1 File of \citet{Oizumi2022PLOS}, this equation is equivalent to
\begin{equation}
\det\left(I-\frac{1}{\lambda}A\right)=0.
\end{equation}
The dominant eigenvalue is characterized as the value for which all first-return pathways to the reference state sum to one. This is a multiregional Euler--Lotka equation: it extends the single-region condition that age-specific fertility and survival sum to one to closed pathways that include interregional migration.

\subsection*{Left Eigenvector and Reproductive Value}

The same argument applies to $A^\top$ for a left eigenvector $vB=v$. With $v(\ell)=1$,
\begin{equation}
v(j)=\lambda^{-1}a_{\ell j}
+\sum_{m=1}^{\infty}\sum_{j_1,\ldots,j_m\neq\ell}
\lambda^{-m-1}a_{\ell j_1}a_{j_1j_2}\cdots a_{j_mj}.
\end{equation}
In the notation of the main text, the right-eigenvector series $\pi_{ij}^{j}(m;r)$ describes descendant destinations from origin $j$, whereas the left-eigenvector series $\pi_{ji}^{*j}(m;r)$ describes the origins whose contributions constitute reproductive value in region $j$. The right eigenvector is the stable distribution of birth regions; the left eigenvector is the reproductive value of each birth region for future birth genealogies. They form a natural dual pair between adjoint operators.

\subsection*{Mean Recurrence Generation From the Dual Pairing}

Proposition B.1 identifies the first moment of the normalized first-return distribution with the normalized dual pairing; the standard Perron perturbation formula then gives the reciprocal diagonal-sensitivity relation.

\medskip
\noindent\textbf{Proposition B.1 (Dual-Pairing Identity).}
Let $\boldsymbol\Psi$ be the Perron-normalized form of an irreducible nonnegative next-generation matrix, obtained by evaluating the series at its Perron root so that $\rho(\boldsymbol\Psi)=1$. Let its positive right and left Perron eigenvectors satisfy
\begin{equation}
\boldsymbol\Psi w=w,\qquad v\boldsymbol\Psi=v.
\end{equation}
For reference region $j$, normalize $\widehat w_i=w_i/w_j$ and $\widehat v_i=v_i/v_j$. Let $\pi_{jj}^{j}(m;r)$ be the weight, defined in the main text, of paths that avoid $j$ at intermediate generations and first return to $j$ in generation $m$. Then
\begin{equation}
\sum_{i=1}^{M}\widehat v_i\widehat w_i
=\frac{vw}{v_jw_j}
=\sum_{m=1}^{\infty}m\pi_{jj}^{j}(m;r).
\label{app:identity1}
\end{equation}

\noindent\textit{Proof.}
We explicitly form the Cauchy product of the left- and right-eigenvector series. For $i\neq j$, define the left and right path sums
\begin{equation}
\begin{aligned}
V_i^{(0)}&:=\psi_{ji}(r),\\
V_i^{(a)}&:=\sum_{h_1,\ldots,h_a\neq j}
\psi_{jh_1}(r)\psi_{h_1h_2}(r)\cdots\psi_{h_ai}(r),\qquad a\geq1,\\
W_i^{(0)}&:=\psi_{ij}(r),\\
W_i^{(b)}&:=\sum_{k_1,\ldots,k_b\neq j}
\psi_{ik_1}(r)\psi_{k_1k_2}(r)\cdots\psi_{k_bj}(r),\qquad b\geq1.
\end{aligned}
\end{equation}
The series representations imply
\begin{equation}
\widehat v_i=\sum_{a=0}^{\infty}V_i^{(a)},
\qquad
\widehat w_i=\sum_{b=0}^{\infty}W_i^{(b)},
\qquad i\neq j,
\end{equation}
and $\widehat v_j=\widehat w_j=1$. All terms are nonnegative, so Tonelli's theorem permits interchange of the sums. Therefore,
\begin{equation}
\begin{aligned}
\sum_{i=1}^{M}\widehat v_i\widehat w_i
&=1+\sum_{i\neq j}
\left(\sum_{a=0}^{\infty}V_i^{(a)}\right)
\left(\sum_{b=0}^{\infty}W_i^{(b)}\right)\\
&=1+\sum_{i\neq j}\sum_{a=0}^{\infty}\sum_{b=0}^{\infty}V_i^{(a)}W_i^{(b)}\\
&=1+\sum_{q=1}^{\infty}\sum_{a=0}^{q-1}\sum_{i\neq j}V_i^{(a)}W_i^{(q-1-a)}.
\end{aligned}
\end{equation}
The last equality sets $q=a+b+1$. For fixed $q$ and $a$, interpreting $i$ as the $(a+1)$th interior state of the loop gives
\begin{equation}
\sum_{i\neq j}V_i^{(a)}W_i^{(q-1-a)}
=\sum_{i_1,\ldots,i_q\neq j}
\psi_{ji_1}(r)\psi_{i_1i_2}(r)\cdots\psi_{i_qj}(r).
\end{equation}
The right side does not depend on $a$. The same first-return loop of length $q+1$ appears $q$ times, once for each choice $a=0,1,\ldots,q-1$ of the interior state at which the left and right paths are joined. Hence
\begin{equation}
\sum_{i=1}^{M}\widehat v_i\widehat w_i
=1+\sum_{q=1}^{\infty}q
\sum_{i_1,\ldots,i_q\neq j}
\psi_{ji_1}(r)\psi_{i_1i_2}(r)\cdots\psi_{i_qj}(r).
\label{app:identity2}
\end{equation}
The reference component of the eigenvector equation provides the first-return normalization
\begin{equation}
1=\sum_{m=1}^{\infty}\pi_{jj}^{j}(m;r).
\label{app:identity3}
\end{equation}
Relabeling the number of interior states in Eq.~\eqref{app:identity2} as $q=m-1$ and using Eq.~\eqref{app:identity3} gives
\begin{equation}
1+\sum_{m=2}^{\infty}(m-1)\pi_{jj}^{j}(m;r)
=\sum_{m=1}^{\infty}m\pi_{jj}^{j}(m;r),
\end{equation}
which proves Eq.~\eqref{app:identity1}. \hfill$\square$

It follows that the mean recurrence generation is
\begin{equation}
E_j[m]:=\sum_{m=1}^{\infty}m\pi_{jj}^{j}(m;r)
=\frac{vw}{v_jw_j},
\label{app:identity4}
\end{equation}
which is independent of arbitrary eigenvector scaling. The standard perturbation formula for a simple Perron root further gives
\begin{equation}
\frac{\partial\rho(\boldsymbol\Psi)}{\partial\psi_{jj}}
=\frac{v_jw_j}{vw}
=\frac{1}{E_j[m]}.
\label{app:identity5}
\end{equation}
Mean recurrence is also the reciprocal sensitivity of the Perron root to the diagonal birth-region loop. This identity provides the mathematical basis for treating mean recurrence as the dual pairing of left and right eigenfunctions (or eigenvectors in finite dimensions).

The probability interpretation applies whenever the path weights are evaluated at the Perron root, because Eq.~\eqref{app:identity3} then holds. In that case, $\{\pi_{jj}^{j}(m;r)\}_{m\geq1}$ is the probability distribution of the first-return generation and Eq.~\eqref{app:identity4} is its expectation. A small $E_j[m]$ indicates dominance by short loops; a large value indicates a long tail of loop weights. We apply this interpretation to the birth-genealogy network after normalization to replacement level; it is not the observed real-time return frequency of individuals.

\end{document}